\documentclass[11pt,a4paper]{article}
\usepackage{german}
\usepackage{times}
\usepackage{amsmath}
\usepackage{epsfig}
\usepackage{a4}
\usepackage[hang,stable]{footmisc}
\newtheorem{Theorem}{Theorem}
\newtheorem{Definition}{Definition}
\newtheorem{Bemerkung}{Bemerkung}

\begin{document}

\title{"Uber die Herkunft der Speziellen Relativit"atstheorie\footnote{%
       Erschienen in: Herbert Hunziker (Hrsg.) \emph{Der jugendliche 
      Einstein und Aarau} (Birkh"auser Verlag, Basel, 2005)}}
\author{Domenico Giulini            \\
        Universit"at Freiburg       \\
        Physikalisches Institut     \\
        Hermann-Herder-Stra"se 3    \\
        79104 Freiburg}
\date{}

\maketitle

\begin{abstract}
\noindent
Um 1905 war die Spezielle Relativit"atstheorie zum Greifen nahe --
sie "`lag in der Luft"'. Doch bedurfte es anscheinend der frischen
und unvoreingenommenen Herangehensweise eines Neulings, um den 
letzten, entscheidenden Schritt zu tun. Ich schildere einige der 
begrifflichen Hintergr"unde und Schwierigkeiten, die Anla"s zur 
Aufstellung dieser Theorie gegeben haben. 
\end{abstract}

\section{Einleitung}
\label{sec:Einleitung}
Die Spezielle Relativit"atstheorie (im folgenden mit "`SRT"' abgek"urzt) 
ist aus dem Versuch entstanden,
die Elektrodynamik bewegter K"orper zu verstehen. Dies war keine 
leichte Aufgabe, denn den Elektromagnetismus dachte man sich gegen 
Ende des 19.\,Jahrhunderts als effektive Beschreibung der Dynamik 
eines alles erf"ullenden Mediums, \emph{"Ather } 
genannt\footnote{Der (naive) "Atherbegriff reicht weit in die Antike 
zur"uck; einen kurzen geschichtlichen Abri"s gibt Alexander von Humboldt im 
\emph{Kosmos} (\cite{Humboldt:Kosmos}, p.\,402-408).}, 
dessen physikalische Natur noch weitgehend ungekl"art war. 
Es bestand die weitverbreitete Meinung, da"s sich die 
Dynamik dieses Mediums letztlich auf die Gesetze der Mechanik 
w"urde zur"uckf"uhren lassen. Weniger klar war, ob auch der "Ather 
letztlich eine atomistische Struktur haben w"urde (Mechanik von 
Massenpunkten) oder ob er bis in die kleinsten Dimensionen hinein 
ein Kontinuum bildete. In jedem Fall galt dieses hypothetische 
Medium als zweite S"aule des damaligen, dualistischen Materiekonzepts,
dessen erste S"aule die in Form von gew"ohnlichen K"orpern 
vorliegende Materie umfa"ste, von der mehrheitlich -- aber nicht ohne 
prominente Gegenstimmen\footnote{Zum Beispiel Max Planck, Ernst Mach und 
Wilhelm Ostwald.} -- angenommen wurde, sie w"urde sich letztlich als 
atomistisch aufgebaut erweisen. Da man sich den "Ather gewichtslos 
dachte, bezeichnete man die gew"ohnliche Materie auch als 
\emph{ponderabel}, also w"agbar. 

Um die Elektrodynamik bewegter K"orper zu verstehen, war es also 
notwendig, die feineren Details der Wechselwirkung zwischen "Ather 
und ponderabler Materie zu verstehen. W"urde der "Ather im Inneren 
bewegter K"orper einfach mitgenommen, oder ist er so fein verteilt, 
da"s er durch die R"aume zwischen den Atomen ungehindert "`hindurchwehen"' 
kann? -- oder w"urde er, als eine Art Kompromi"s dieser M"oglichkeiten, 
vielleicht nur teilweise an deren Bewegung teilnehmen? Klar war, da"s 
der "Ather durch die K"orper nicht einfach verdr"angt w"urde, denn der 
"Ather war Sitz elektromagnetischer Felder und des Lichtes, die sich 
eben auch in Festk"orpern ausbreiten k"onnen, wie man etwa an einem 
Lichtstrahl im Glas eindr"ucklich sieht. 

Viele komplizierte Theorien des "Athers wurden erdacht, doch keine
war wirklich befriedigend. Die Schwierigkeit war, da"s sich trotz 
feinsinnig konzipierter Experimente eine Bewegung relativ zum "Ather 
nicht feststellen lie"s, was der Intuition zuwiderlief: Wenn 
elektromagnetische Wellen ein periodischer Ausbreitungsvorgang 
\emph{im "Ather} sind, dann mu"s sich eine Bewegung relativ zum "Ather 
doch an diesen Wellen messen lassen -- so schlo"s man aus Gr"unden 
der Analogie zu Wasser- oder Schallwellen. Die Schwierigkeiten schienen 
un"uberwindbar, bis Einstein mit seiner Speziellen Relativit"atstheorie
den Gordischen Knoten auf eine das physikalische Publikum "uberraschende 
Weise zerschlug. Die begriffliche Pointe seiner L"osung ist, da"s es 
zur Beschreibung der Ph"anomene der Vorstellung des "Athers "uberhaupt 
nicht bedarf. 

Meist wird diese Pointe in der rigoroseren Form dargestellt, da"s Einstein 
den "Ather regelrecht \emph{abgeschafft} hat. Dies ist strenggenommen aber
nicht zwingend, sondern wird nur durch Ockhams Rasiermesser\footnote{%
Als "`Ockhams Rasiermesser"' bezeichnet man das auf William von Ockham 
(1287-1347) zur"uckgehende erkenntnistheoretische 
Prinzip, nach dem von zwei Theorien mit dem gleichen, die Ph"anomene 
betreffenden Erkl"arungswert diejenige zu bevorzugen ist, die mit der 
geringeren Anzahl von Hypothesen auskommt. Insbesondere sind als 
"uberfl"ussige Hypothesen solche zu betrachten, die ontologische 
Entit"aten setzen, die keinerlei Auswirkung auf die Ph"anomene haben, 
wie hier der "Ather. Man sollte jedoch auch betonen, da"s eine 
allzu kurzsichtige Anwendung dieses Prinzips insofern auch sch"adlich
sein kann, als damit kurzerhand Strukturelemente verworfen werden, 
die sich m"oglicherweise bei der zuk"unftigen Entwicklung der Theorie 
und ihrer Fragestellungen als fruchtbar erwiesen h"atten. 
Hier ist also vor allem "`die gute Nase"' des Forschers gefragt, der 
sich Einstein einmal, nebst seiner "`maultierhaften Starrn"ackigkeit"', 
als einzig bemerkenswerte Eigenschaft r"uhmte (\cite{Seelig:HelleZeit}, 
p.\,72.)}
nahegelegt. Was Einstein wirklich zeigte, ist, da"s man dem "Ather
keine kinematischen Bewegungsgr"o"sen zuschreiben 
darf (vergleiche dazu Einsteins klare Diskussion 
in~\cite{Einstein:Aether}). Man wird jedoch zumindest sagen k"onnen, 
da"s ein so, jeder kinematischer Bewegungsgr"o"sen entbehrendes Etwas, 
kaum noch mit einer \emph{substantiellen} "Athervorstellung vertr"aglich 
ist, so da"s es nat"urlicher erscheint, diesen Begriff ganz zu vermeiden. 
Da"s dies widerspruchsfrei m"oglich ist, zeigt die SRT.

Die zuletzt gemachten Bemerkungen wurden deshalb so betont, weil 
sie f"ur die begriffliche Entwicklung der Physik des 20.\,Jahrhunderts 
besonders wichtig sind, insbesondere f"ur den heutigen Begriff des 
\emph{Feldes}, der sowohl in klassischen wie in Quantentheorien eine 
zentrale Rolle spielt. Diese Felder sind definiert in Raum und 
Zeit, \emph{ohne} da"s damit die Vorstellung einer substantiellen 
Tr"agersubstanz verbunden w"are. Dieses abstrakte Konzept gibt es in 
dieser Form in der Physik erst seit Aufstellung der SRT. 
Heutige Lehrb"ucher vermitteln kaum noch eine Ahnung davon, welcher 
begrifflichen (und psychologischen) Anstrengungen es bedurfte, um 
den Begriff des Feldes von der gedanklichen Bindung an ein 
substantielles Tr"agermedium zu l"osen und so als selbst"andiges 
physikalisches Konzept zu etablieren.

\section{Relativit"atsprinzip und Tr"agheitsgesetz}
\label{sec:RelPrinzTraeg}
Das Relativit"atsprinzip ist eines der Grundpfeiler der klassischen 
Mechanik. Seine erste und vielleicht auch sch"onste Formulierung stammt 
von Galilei. In seinem ber"uhmtesten Werk, dem "`Dialog "uber die beiden 
haupts"achlichsten Weltsysteme"' aus dem Jahre 1632 -- unter Kennern 
meist nur kurz \emph{der Dialog} genannt -- legt er seinem alter ego, 
Salviati\footnote{Filipo Salviati (1582-1614) entstammte einer 
Florentiner Patrizierfamilie und war bis zu seinem Tode ein guter 
Freund Galileis. Unter den drei Galileischen Dialogfiguren
Salviati, Sagredo und Simplicio spielt er die Rolle des 
belehrenden Akademikers, der die "Uberzeugungen Galileis zum 
Ausdruck bringt.}, folgende Worte in den Mund:   
\begin{quote}
"`Schlie"st Euch in Gesellschaft eines Freundes in einen m"oglichst 
gro"sen Raum unter dem Deck eines gro"sen Schiffes ein. Verschafft 
Euch dort M"ucken, Schmetterlinge und anderes fliegendes Getier;
sorgt auch f"ur ein Gef"a"s mit Wasser und kleinen Fischen darin;
h"angt ferner oben einen kleinen Eimer auf, welcher tropfenweise 
Wasser in ein zweites enghalsiges darunter gestelltes Gef"a"s tropfen
l"a"st. Beobachtet nun sorgf"altig, solange das Schiff stille steht, 
wie die fliegenden Tierchen mit der n"amlichen Geschwindigkeit nach 
allen Seiten des Zimmers fliegen. Man wird sehen, wie die Fische ohne 
irgend welchen Unterschied nach allen Richtungen schwimmen; 
die fallenden Tropfen werden alle in das untergestellte Gef"a"s flie"sen. 
Wenn Ihr Eurem Gef"ahrten einen Gegenstand zuwerft, so braucht Ihr 
nicht kr"aftiger nach der einen als nach der anderen Richtung zu werfen,
vorausgesetzt, da"s es sich um gleiche Entfernungen handelt. Wenn Ihr, 
wie man sagt, mit gleichen F"u"sen einen Sprung macht, werdet Ihr nach 
jeder Richtung hin gleichweit gelangen. [...]
Nun la"st das Schiff mit jeder beliebigen Geschwindigkeit sich bewegen:
Ihr werdet -- wenn nur die Bewegung gleichf"ormig ist und nicht hier- 
und dorthin schwankend -- bei allen genannten Erscheinungen nicht die 
geringste Ver"anderung eintreten sehen. Aus keiner derselben werdet Ihr 
entnehmen k"onnen, ob das Schiff f"ahrt oder stille steht."'
\end{quote}
Etwas prosaischer ausgedr"uckt wird hier behauptet, da"s man durch 
keinerlei Experimente unter Deck feststellen kann, ob sich das Schiff 
geradlinig-gleichf"ormig "uber das Wasser bewegt oder nicht.\footnote{%
Strenggenommen spricht Galilei das Tr"agheitsprinzip immer nur f"ur 
horizontale Bewegungen auf der Erdoberfl"ache aus, weil er zwar von 
allen horizontal wirkenden St"oreinfl"ussen, nicht jedoch vom 
vertikal gerichteten Schwerefeld der Erde abstrahiert. Trotzdem ist 
der Abstraktionsgedanke, der die wesentliche Leistung bei der 
Formulierung dieses Prinzips ausmacht, bei Galilei klar vorhanden 
(in 2 von 3 Richtungen), so da"s es gerechtfertigt erscheint,
das allgemeine Prinzip als eine \emph{seiner} gro"sen Leistungen zu 
bezeichnen~\cite{Fierz:Mechanik}.}

Man beachte -- um bei Galileis Beispiel zu bleiben --, da"s \emph{alle} 
Komponenten des betrachteten physikalischen Systems gleicherma"sen 
der translatorischen Bewegung des Schiffes unterworfen werden. Dazu 
z"ahlt insbesondere auch die von den Schiffsw"anden eingeschlossene 
Luft, die man sich auch bei fahrendem Schiff relativ zu diesem ruhend 
denken soll. Aus diesem Grund sagt Galilei zu Beginn, da"s man 
"`unter Deck"' gehen solle. W"urden die Luken des Schiffes ge"offnet, 
so da"s die Luft hindurchstreichen kann, so w"urden die Schmetterlinge 
relativ zum Schiff nat"urlich keineswegs mehr eine nach allen Richtungen 
gleiche Geschwindigkeitsverteilung zeigen, sondern sich bevorzugt 
in Richtung des durchstr"omenden Windes bewegen. Wir werden auf 
dieses Bild bei der Diskussion des Relativit"atsprinzips in der 
Elektrodynamik zur"uckkommen. 

Ebenso ist hervorzuheben, da"s alle von Galilei erw"ahnten Ph"anomene 
prim"ar mechanischer Natur sind, d.h. durch die Gesetze der Mechanik 
im Prinzip beschrieben werden k"onnen. Die moderne und abstraktere 
Version des hier ausgesprochenen mechanischen Relativit"atsprinzips 
lautet dann so: 
\par\medskip
\noindent
\textbf{Mechanisches Relativit"atsprinzip.}\emph{
Zwei identische, abgeschlossene physikalische Systeme, die sich 
relativ zueinander in gleichf"ormig geradliniger Bewegung 
befinden, sind hinsichtlich der an den Einzelsystemen mechanisch 
me"sbaren Ph"anomene ununterscheidbar.}
\medskip

\noindent
Mathematisch gesprochen ist "`relativ zueinander geradlinig gleichf"ormig 
bewegt"' eine "Aquivalenzrelation auf der Menge abgeschlossener 
physikalischer Systeme. Das mechanische Relativit"atsprinzip besagt 
dann, da"s Elemente einer "Aquivalenzklasse \emph{mechanisch} 
ununterscheidbar sind. Die SRT entstand aus der Frage, ob sie 
vielleicht \emph{elektrodynamisch} unterscheidbar sein w"urden. 
Doch vorerst verbleiben wir noch bei der Mechanik. 

Beziehen wir die eben eingef"uhrte "Aquivalenzrelation auf physikalische 
Bezugssysteme, so besagt das Tr"agheitsgesetz der Mechanik, da"s unter den 
"Aquivalenzklassen gleichf"ormig geradlinig zueinander bewegter 
Bezugssysteme eine Klasse ausgezeichnet ist: die Klasse der 
\emph{Inertialsysteme}. Bez"uglich diesen -- und nur diesen -- gilt die
Newtonsche Gleichung\footnote{Newton schrieb eigentlich (in moderner
Schreibweise) $\vec F=\dot{\vec p}$, wobei $\vec p$ der Impuls ist 
und ein Punkt dar"uber die zeitliche Ableitung bezeichnet. 
Erst durch Setzen von $\vec p=m\vec v$, mit $m=$ Masse und 
$\vec v=$ Geschwindigkeit wird daraus Gleichung (\ref{eq:Newton}), 
\emph{falls} $m$ konstant ist. Letzteres mu"s nicht immer gelten; man 
denke etwa an eine Rakete, die durch Materialaussto"s st"andig an 
Masse verliert.}
\begin{equation}
\label{eq:Newton}
\vec F=m\vec a\,.
\end{equation}
Insbesondere folgt, da"s bez"uglich solcher Systeme ein kr"aftefreier
K"orper unbeschleunigt ist, d.h. sich geradlinig und gleichf"ormig 
bewegt. Dies ist der Inhalt des sogenannten Tr"agheitsgesetzes: es 
besagt die \emph{Existenz} -- und nicht mehr! -- ausgezeichneter, 
raumzeitlicher Bezugssysteme, denen gegen"uber kr"aftefreie K"orper -- 
meist als Massenpunkte idealisiert, um mit ihnen eine eindeutige Bahn 
assoziieren zu k"onnen -- stets gleichf"ormig geradlinig bewegt sind. 
Wie bereits betont, macht das Tr"agheitsgesetz keine konkreten Angaben 
"uber eine physikalische Realisierung eines Inertialsystems. Diese 
Unbestimmtheit (die in Lehrb"uchern meist "uberhaupt nicht 
problematisiert wird) hat Einstein einmal so kommentiert 
(die Hervorhebungen sind die Einsteins; siehe \cite{Einstein-CW}, 
Band\,7, Dokument\,49): 
\begin{quote}
,,Voneinander hinreichend entfernte, materielle Punkte bewegen sich 
geradlinig gleichf"ormig -- \emph{vorausgesetzt, da"s man die Bewegung 
auf ein passend bewegtes Koordinatensystem bezieht und da"s man die 
Zeit passend definiert.} Wer empfindet nicht das Peinliche einer 
solchen Formulierung? Den Nachsatz weglassen aber bedeutet
eine Unredlichkeit.``
\end{quote}

Oft wird angenommen, da"s Inertialsysteme durch ausgezeichnete 
astrophysikalische Bezugssysteme realisiert werden, etwa dem 
Hintergrund der galaktischen ,,Fixsterne''. Doch ist das bestenfalls 
eine sehr gute Approximation, denn neben den irregul"aren 
Pekuliargeschwindigkeiten (die man im Prinzip herausmitteln kann) 
f"uhren alle Sterne unserer Milchstra"se kollektiv eine globale 
Rotationsbewegung aus.\footnote{Unser Sonnensystem bewegt 
sich in etwa 225 Millionen Jahren einmal um das galaktische Zentrum. 
Von uns aus gesehen f"uhrt das zu einer kollektiven Bewegung der 
galaktischen Sterne gegen"uber dem Hintergrund der entferntesten 
extragalaktischen Objekte (Quasare) von etwa $5$ Millibogensekunden 
pro Jahr. Das ist zwar extrem wenig, aber nicht au"serhalb 
gegenw"artiger Me"stechnologien. So mi"st das Gravity-Probe-B-Experiment
\cite{GPB} das Tr"agheitsverhalten von Kreiseln mit 10-fach besserer 
Aufl"osung.}
F"ur astronomische Zwecke verwendet man seit 1998 das \emph{International 
Celestial Reference System}~\cite{ICRF}, das sich an "uber 600 
extragalaktischen Radioquellen orientiert, deren Positionen mit 
einem weltweiten Verbund von Radioteleskopen (VLBI) sehr genau 
bestimmt werden. 

Nat"urlich haftet solchen extrinsischen Identifikationen immer eine 
gewisse Willk"ur an, jedenfalls so lange, wie nicht auch aus 
fundamentalen dynamischen Prinzipien heraus verstanden ist, warum 
gerade das eine, an real existierenden K"orpern festgemachte 
Bezugssystem ein Inertialsystem sein soll. Deshalb soll noch ein 
intrinsisches Verfahren zur Konstruktion von Inertialsystemen erw"ahnt 
werden, das auf Ludwig Lange zur"uckgeht~\cite{Lange:1885}. Von ihm 
stammt "ubrigens auch  die Terminologie \emph{Inertialsystem} und 
\emph{Inertialzeitskala}. Das Anliegen Langes war es, den definitorischen 
Teil des Tr"agheitsgesetzes von seinem nichttrivialen physikalischen 
Gehalt sauber zu trennen, um so letzteren "uberhaupt erst richtig sichtbar 
zu machen. Das Tr"agheitsgesetz zerf"allt so in zwei Definitionen und 
zwei Theoreme (physikalisch, nicht mathematisch verstanden) wie folgt:
\begin{Definition}
Ein Koordinatensystem hei"st {\textbf{Inertialsystem}}, wenn sich ihm 
gegen"uber drei von einem gemeinsamen Punkt nicht koplanar 
weggeschleuderte kr"aftefreie Massenpunkte auf Geraden bewegen. 
\end{Definition}
\begin{Theorem}
Bez"uglich eines so festgelegten Inertialsystems ist die Bahn 
\textbf{jedes} kr"aftefreien Massenpunktes eine Gerade.
\end{Theorem}
\begin{Bemerkung}
Durch die ersten drei kr"aftefrei bewegten Massenpunkte wird das 
Inertialsystem "uberhaupt erst operational definiert und damit die 
Bedeutung von ,,gerade''. Erst mit vier und mehr kr"aftefreien 
Massenpunkten ist sein physikalischer Gehalt "uberpr"ufbar.
\end{Bemerkung}
\begin{Definition}
Eine Zeitskala hei"st \textbf{Inertialzeitskala}, wenn bez"uglich ihr ein 
kr"aftefreier Massenpunkt in gleichen Zeiten gleiche Strecken 
zur"ucklegt.
\end{Definition}
\begin{Theorem}
Bez"uglich einer so festgelegten Inertialzeitskala wird die Bahn  
\textbf{jedes} kr"aftefreien Massenpunktes gleichf"ormig durchlaufen. 
\end{Theorem}
\begin{Bemerkung}
Hier wird durch den ersten kr"aftefrei bewegten Massenpunkt die 
Zeitskala definiert und damit die Bedeutung von ,,gleichf"ormig''. 
Dabei ist vorausgesetzt, da"s ein r"aumliches Abstandsma"s 
existiert. Erst anhand eines zweiten, dritten etc. kr"aftefreien 
Massenpunktes kann dann die Gleichf"ormigkeit ihrer Bewegung 
"uberpr"uft werden. 
\end{Bemerkung}

Die hier wiedergegebenen Theoreme stellen eine reine Erfahrungstatsache 
dar. Ob und gegebenenfalls wie die durch sie ausgedr"uckte Auszeichnung 
der Inertialsysteme und Inertialzeitskalen aus der Menge aller 
Bezugssysteme und Zeitskalen physikalisch kausal (d.h. als Resultat 
dynamischer Wirkungen) zu verstehen ist, bleibt hier v"ollig offen. 
Newton setzt an diese Stelle gewisse metaphysische Konstrukte, 
genannt den \emph{Absoluten Raum} und die \emph{Wahre Zeit}, deren Existenz
nach Newtons Auffassung gerade durch die G"ultigkeit des 
Tr"agheitsgesetzes in Evidenz gesetzt wird. An dieser Existenz 
absoluter Strukturen\footnote{Das Wort "`absolut"' soll in 
diesem Zusammenhang andeuten, da"s die betreffende Struktur nicht 
selbst Gegenstand dynamischer Ver"anderungen ist, obwohl von ihr 
dynamische Wirkungen ausgehen. So ist in der Newtonschen Mechanik 
der Absolute Raum (und nicht irgendwelche K"orper) die gedachte 
Ursache f"ur das Auftreten von Tr"agheitskr"aften, wie der 
Zentrifugal- oder Corioliskraft. Eine r"uckwirkende Kraft und 
damit dynamische Beeinflussung realer K"orper auf den absoluten Raum 
gibt es aber nicht, also auch kein allgemeing"ultiges "`actio gleich 
reactio"'. Die "Uberwindung dieses erkenntnistheoretisch 
unbefriedigenden Aspekts des Newtonschen Theoriengeb"audes war eine 
der Hauptmotivationsstr"ange, die Einstein zur Aufstellung der 
Allgemeinen Relativit"atstheorie trieben.}
wird auch die Spezielle Relativit"atstheorie im Prinzip nur sehr wenig 
"andern. Lediglich die absolute Unterscheidung r"aumlicher 
(Bezugssystem) und zeitlicher (Zeitskala) Aspekte wird zugunsten eines
einheitlichen Begriffes des raum-zeitlichen Bezugssystems aufgegeben. 
Die absolute Auszeichnung raum-zeitlicher Inertialsysteme bleibt aber 
weiterhin bestehen. 

Ein Bezugssystem wird durch Angabe von vier Koordinatenfunktionen 
$(t,\vec x)$ auf der Menge der Ereignisse (der Raum-Zeit) gegeben. 
Damit wird jedem Ereignis in umkehrbar eindeutiger Weise ein Quadrupel 
reeller Zahlen zugeordnet. Diese haben wir bereits so geschrieben, da"s
$t$ als Zeit- und $\vec x$ als Ortskoordinate interpretiert wird. 
Ist ein Inertialsystem $(t,\vec x)$ gegeben, bez"uglich dem die 
Newtonschen Bewegungsgleichungen (\ref{eq:Newton}) gelten, so 
gelten sie auch bez"uglich jedem anderen Inertialsystem $(t',\vec x')$, 
das sich von ersterem durch folgende Transformationen unterscheidet
\footnote{Wir setzen hier implizit voraus, da"s die Kr"afte $\vec F$
zwischen den Punktteilchen (aus denen wir alles aufgebaut denken) 
nur von den momentanen Abst"anden der Teilchen abh"angen.}: 
1)~konstante Translationen der Zeit- und Ortskoordinaten, 
2)~r"aumliche Drehungen und 
3)~Geschwindigkeitstransformationen. Letztere haben die Form 

\begin{subequations}
\label{eq:GalileiTrans}
\begin{alignat}{2}
\label{eq:GalileiTrans1}
& \vec x'&&\,=\,\vec x-\vec vt\,,\\
\label{eq:GalileiTrans2}
& t'&&\,=\,t\,.
\end{alignat}
\end{subequations}
und werden auch (eigentliche) Galilei-Transformationen genannt. 

Betrachtet man einen Gegenstand, der sich im gestrichenen System mit 
der konstanten Geschwindigkeit $\vec w'$ bewegt, dort also der Gleichung 
$\vec x'(t')=\vec w'\,t'$ gen"ugt, so bewegt er sich bez"uglich 
des ungestrichenen Systems gem"a"s (\ref{eq:GalileiTrans}) nach 
der Gleichung $\vec x(t)=\vec w\,t$, also mit der Geschwindigkeit 
$\vec w$, wobei  
\begin{equation}
\label{eq:GeschAddGal}
\vec w=\vec w'+\vec v\,.
\end{equation} 
Dies ist das klassische Gesetz, nach dem die \emph{physikalische} 
Operation der Komposition von Geschwindigkeiten durch die 
\emph{mathematische} Operation der Addition von Vektoren abgebildet 
wird. Diese "Ubersetzungsvorschrift ist keineswegs so evident wie 
man vielleicht meinen k"onnte. Sie wird sich in der SRT auch als 
falsch erweisen. 

An dieser Stelle will ich auch einen Punkt ansprechen, der eine meiner 
Meinung nach potentiell irref"uhrende Sprachregelung angeht: 
Solange man die Invarianz eines dynamischen Gesetzes unter 
Galilei-Transformationen (\ref{eq:GalileiTrans}) als mathematischen Ausdruck 
des (physikalisch definierten) Relativit"atsprinzips ansieht, ist das 
betreffende Gesetz auch "`relativistisch"' zu nennen. In \emph{diesem} 
Sinne ist insbesondere die klassische Mechanik voll und ganz 
"`relativistisch"'. Trotzdem wird sie im Vergleich zur Mechanik der 
SRT "'nichtrelativistisch"' genannt (worin eben die genannte 
Irref"uhrung liegt), weil man seit Aufstellung der SRT zu 
der "Uberzeugung gekommen ist, da"s das Relativit"atsprinzip korrekt 
nicht durch die Galilei-Transformationen (\ref{eq:GalileiTrans}), 
sondern durch die Lorentz-Transformationen gegeben ist. 
Entsprechend wird im heutigen Sprachgebrauch ein dynamisches Gesetz 
nur dann "`relativistisch"' genannt, wenn es unter 
Lorentz-Transformationen invariant ist, was auf die klassische 
Mechanik eben nicht zutrifft. 

Zum Schlu"s dieses Abschnitts m"ochte ich die Gelegenheit nutzen, die 
Lorentz-Transformationen einmal aufzuschreiben. 
Bezeichnen wir mit $\vec x_{\Vert}$ und $\vec x_{\perp}$ die
Orthogonalprojektionen von $\vec x$ parallel bzw. senkrecht zu 
$\vec v$, so ist  
\begin{subequations}
\label{eq:LorentzTrans}
\begin{alignat}{2}
\label{eq:LorentzTrans1}
& \vec x'&&\,=\,\gamma\,(\vec x_{\Vert}-\vec vt) + \vec x_{\perp}
=\,\vec x-\vec vt+(\gamma-1)(\vec x_{\Vert}-\vec vt)\,,\\ 
\label{eq:LorentzTrans2}
& t'&&\,=\,\gamma\,(t-\vec x\cdot\vec v/c^2)\,,\\
\label{eq:LorentzTrans3}
\text{wobei}\qquad  
& \gamma&&\,=\,1/\sqrt{1-v^2/c^2}\,.
\end{alignat}
\end{subequations}

\section{Das dualistische Materiekonzept des 19. Jahrhunderts}
\label{sec:Materiekonzept}
Die Newtonsche Mechanik basiert wesentlich auf dem Konzept des 
\emph{Punktteilchens}. Diesem kann man einen scharfen Ort $\vec x$ zuweisen 
und damit auch eine scharf definierte Bahn:  $t\mapsto \vec x(t)$. 
Kennt man die Verteilung der Kraft als Funktion des Ortes und der Zeit, 
so kann man aus (\ref{eq:Newton}) die Bahn nach Vorgabe eines Anfangsortes 
und einer Anfangsgeschwindigkeit eindeutig bestimmen. Kennt man 
umgekehrt die Bahn, so kann man die Kr"afte errechnen, die an den 
verschiedenen Orten der Teilchenbahn zu den entsprechenden Zeiten 
gewirkt haben.

Mit Hilfe dieses Konzepts idealer Punktteilchen gelingt es, die 
Bewegung komplizierter zusammenh"angender Konfigurationen solcher 
Teilchen auf die Bewegungsgesetze dieser Teilchen zur"uckzuf"uhren, 
sofern einfache Annahmen "uber die zwischen den Punktteilchen 
wirkenden Kr"afte gemacht werden. Etwas idealisierend kann man sich 
z.B. Kr"afte vorstellen, die die einzelnen Teilchen in einer 
starren Konfiguration halten, wo also die relativen Positionen 
der Teilchen untereinander immer dieselben sind. Man kommt so zum 
Begriff des (idealen) \emph{starren K"orpers}, dessen Konfigurationen 
im Raum durch die Angabe von nur sechs Zahlen vollst"andig 
charakterisiert werden k"onnen, n"amlich drei Koordinaten zur 
Festlegung eines beliebigen seiner Punkte -- etwa des Schwerpunktes -- 
und drei zur Festlegung seiner dann noch verbleibenden 
Drehfreiheiten um diesen Punkt. Nat"urlich handelt es sich dabei 
um eine N"aherung, die nur solange erlaubt ist, wie die von au"sen 
am Gesamtsystem angreifenden Kr"afte (etwa Gravitationskr"afte) sehr 
klein sind im Vergleich zu den inneren Kr"aften, die die Teilchen 
gegenseitig fixieren (etwa elektrostatische Kr"afte). Allgemein 
erweist sich dieses reduktionistische Programm als "uberaus 
fruchtbar und f"uhrt zu einer fast un"ubersehbaren F"ulle von 
Anwendungen. Die Dynamik des Fahrrades f"allt ebenso darunter wie 
die Rotation der Erde oder die Bewegung der Planeten gegeneinander. 
Wenn man sich also fragt, auf welche materiellen Entit"aten die 
Newtonsche Mechanik prinzipiell angewendet werden kann, so lautet 
die Antwort, da"s daf"ur alles in Frage kommt, was man sich aus 
diesen Punktteilchen im Prinzip aufgebaut denken kann. 
Stellte man sich auf den Standpunkt eines (naiven) materiellen 
Atomismus, so k"ame man schlie"slich sogar zu der Vermutung, da"s 
sich letztlich \emph{alle} physikalischen Vorg"ange auf einfache 
mechanische Grundgesetze zur"uckf"uhren lassen w"urden. 

Wirklich alle? Schon Newton waren die optischen Erscheinungen 
wohl vertraut. "Uber die Natur des Lichtes und die Gesetze seiner 
Ausbreitung hat auch er spekuliert (in seiner "`Optick"' 
aus dem Jahre~1704), ohne jedoch daf"ur ein Lehrgeb"aude, vergleichbar 
der Mechanik, gr"unden zu k"onnen. Tats"achlich nahm Newton an, 
da"s auch Licht aus kleinsten Teilchen best"unde, die durch Kr"afte, 
wie die Gravitationskraft, Einwirkungen erfahren k"onnen. 
Diese Teilchenvorstellung des Lichtes verschwand aber vollends
zugunsten einer konkurrierenden Vorstellung von Licht als Welle, 
als zu Beginn des 19.\,Jahrhunderts Thomas Young (1773-1829) 
experimentell die Interferenzf"ahigkeit von Licht nachwies, die 
der Teilchenvorstellung kra"s widerspricht. Doch wenn Licht eine 
Welle ist, also ein sich ausbreitender periodischer Schwingungsvorgang, 
so liegt die Frage nahe, \emph{was} da schwingt. Analog der 
Wasserwelle auf der Oberfl"ache eines ruhigen Sees, in der die 
Wasserteilchen mit vertikaler Amplitude im Raum schwingen, m"u"ste 
auch Licht den Schwingungen eines gewissen hypothetischen Mediums 
entsprechen: dem bereits erw"ahnten "Ather. Dieser m"u"ste auch in 
alles eindringen k"onnen, in dem Licht sich fortpflanzt, z.B. in Glas, 
was immerhin eine nicht unerhebliche Dichtigkeit aufweist. Weiterhin 
war schon lange durch die Messungen des d"anischen Astronomen Ole 
R{\o}mer (1644-1710) aus den Jahren 1672-76 bekannt, da"s die 
Lichtgeschwindigkeit einen zwar endlichen aber dennoch extrem hohen 
Wert besitzt, den R{\o}mer damals noch mit 220-tausend Kilometer pro 
Sekunde angab, was etwa 3/4 des heute exakten Wertes 
betr"agt:\footnote{Der Wert ist heute \emph{exakt}, weil die 
Conf\'erence G\'en\'erale des Poids et Mesures im Jahre 1983 die 
Definition des Meters wie folgt festgelegt 
hat: "`Ein Meter ist die Distanz, die das Licht im Vakuum in einer 
Zeit von $1/299\,792\,458$ Sekunden zur"ucklegt"'. Die Sekunde ist    
dabei "uber die Frequenz der Hyperfeinaufspaltung des Grundzustandes 
des Atoms C"asium 133 definiert.} 
\begin{equation}
\label{eq:Lichtgeschwindigkeit}
c=299\,792\,458\ \, m\cdot s^{-1}\,.
\end{equation}
Die ersten terrestrischen Pr"azisionssmesungen der Lichtgeschwindigkeit 
erfolgten 1849-50 durch Fizeau und Foucault, mit Werten, die dem Wert
(\ref{eq:Lichtgeschwindigkeit}) bis auf einige Promille 
nahe kamen.\footnote{Das erste, Fizeausche Experiment, das noch einen 
um $4{,}7\%$ zu gro"sen Wert ergab, ist kurz in Humboldts \emph{Kosmos}
geschildert (\cite{Humboldt:Kosmos}, p.\,428-429).}

Aus der extremen H"ohe der Lichtgeschwindigkeit wird sofort klar, da"s 
zumindest die naive Analogie der Lichtwelle zu einer elastischen 
Verformungswelle eines herk"ommlichen Materials dem "Ather ganz 
ungew"ohnliche Eigenschaften zuweist. Zum Beispiel ist die 
Ausbreitungsgeschwindigkeit in einem elastischen Medium proportional 
zur Wurzel aus dem Quotienten zwischen der Festigkeit und der Dichte. 
Also mu"ste der "Ather bei relativ geringer Dichte eine geradezu 
phantastische Festigkeit besitzen. Setzt man voraus, da"s sich seine 
Festigkeit nach Eintritt in einen K"orper nicht "andert, so ergibt 
sich auch, da"s seine Dichte innerhalb des K"orpers in der Regel 
\emph{gr"o"ser} sein mu"s als au"serhalb, denn die Lichtgeschwindigkeit 
ist innerhalb in der Regel kleiner als au"serhalb. Erschwerend kommt 
hinzu, da"s wegen der Transversalit"at der Lichtwelle der "Ather eher 
einem Festk"orper denn einer Fl"ussigkeit gleichen m"u"ste. 
(Transversalwellen gibt es nur in Medien, die elastische Scherkr"afte 
unterst"utzen k"onnen.) Offensichtlich passen all diese Eigenschaften 
nicht recht zusammen.\footnote{Eine sch"one 
Diskussion dieser Schwierigkeiten auf moderatem technischen Niveau 
enth"alt~\cite{Born:RT}.} 

Trotzdem hielt man aber an dem Konzept eines "Athers fest -- ohne ihn 
freilich physikalisch zu verstehen --, denn ohne ihn erschien nicht 
nur die Wellentheorie des Lichtes ohne physikalische Basis, auch die 
"Ubertragung von Kraftwirkungen "uber mitunter gro"se r"aumliche 
Distanzen schien nicht verst"andlich, wenn nicht ein vermittelndes 
Medium angenommen wurde, das den Kraft"ubertrag physikalisch 
bewerkstelligte. Das entsprach "ubrigens auch Newtons Vorstellung, 
trotz seines scheinbar auf instantaner Wechselwirkung beruhenden 
Gravitationsgesetzes. In einem wunderbaren Brief aus dem Jahre 1693 
an Bentley bezeichnete er die Annahme einer unvermittelten 
und instantanen Kraftausbreitung durch den leeren Raum sogar als 
"`Absurdit"at, auf die niemand verfallen k"onne, der in 
philosophischen Dingen auch nur einigerma"sen kompetent denken 
k"onne"' (\cite{Newton:Briefe}, Brief\,Nr.\,406, p.\,254.).

Ganz grob kann man die "Athertheorien des 19.\,Jahrhunderts in zwei 
Kategorien einteilen, die mit den Namen Fresnels und Lorentz' einerseits 
und den Namen Stokes' und Hertz' andererseits verbunden sind. Der 
wesentliche Unterschied beider besteht f"ur uns in dem Ma"s, mit dem 
der "Ather an der Bewegung der Materie teilnimmt ("`mitgeschleppt"' 
wird). Man spricht hier auch von "`Mitf"uhrung"'. Dabei stehen die 
Namen Fresnel und Lorentz f"ur "`nur teilweise"' bzw. "`gar keine"'
Mitf"uhrung, w"ahrend mit den Namen Stokes und Hertz eine 
"`im wesentlichen vollst"andige"' bzw. "`strikt vollst"andige"' 
Mitf"uhrung assoziiert wird. Das Hauptargument gegen eine strikte 
Mitf"uhrung, so z.B. auch durch die Atmosph"are der Erde, war das 
Ph"anomen der stellaren Aberration, weshalb die meisten Forscher 
den Ansatz Fresnels bevorzugten, auf den wir noch kurz eingehen 
wollen.      

Der erste Versuch, eine Bewegung der Erde relativ zu einem global 
bevorzugten Bezugssystem (hier: "Atherruhesystem) durch optische 
Experimente nachzuweisen, wurde im Jahre 1810 durch den Franzosen 
Fran{\c c}ois Arago (1786-1853) gemacht~\cite{Arago:1810}.\footnote{%
Arago ging urspr"unglich wohl von einer Teilchentheorie des Lichtes 
aus, in der das bevorzugte Bezugssystem Newtons Absoluter Raum w"are. 
Es ist aber f"ur die Wellen-"Ather-Theorie genauso relevant. Er 
untersuchte mit Hilfe eines Prismas, ob der Brechungsindex (vgl.
Fu"snote\,\ref{foot:Brechungsindex}) des Sternenlichtes von der 
Beobachtungsrichtung, d.h. der Richtung relativ zur Geschwindigkeit 
gegen"uber dem global bevorzugten Bezugssystem abh"angt, fand aber 
keinen diesbez"uglichen Effekt. Eine moderne Diskussion 
gibt~\cite{Ferraro:2005}. Das Experiment Aragos ist auch in 
Humboldts \emph{Kosmos} erw"ahnt (\cite{Humboldt:Kosmos}, p.\,429).}
Den negativen Ausgang dieses Experiments erkl"arte Augustin Fresnel 
(1788-1827) im Jahre 1818 dadurch, da"s das Licht im Inneren eines 
mit der Geschwindigkeit $v$ (relativ zum umgebenden "Ather) bewegten, 
durchsichtigen  K"orpers vom 
Brechungsindex\footnote{\label{foot:Brechungsindex} Der \emph{Brechungsindex} 
$n$ eines Materials gibt das Verh"altnis der Lichtgeschwindigkeit 
(genauer: Phasengeschwindigkeit) $c$ im Vakuum (hier: freien "Ather) zur 
Lichtgeschwindigkeit $c'$ innerhalb des K"orpers an: $n=c/c'$.} $n$ 
einen Geschwindigkeitszuwachs um den Bruchteil
\begin{equation}
\label{eq:FresnelKoeff}
f=1-n^{-2}
\end{equation}
der K"orpergeschwindigkeit in deren Richtung erf"ahrt~\cite{Fresnel:1818}. 
Dies deutete er einfach so, 
da"s der "Ather im Inneren des bewegten K"orpers eben genau um den 
durch (\ref{eq:FresnelKoeff}) gegebenen Bruchteil mitgef"uhrt w"urde. 
Bis heute nennt man $f$ den \emph{Fresnelschen Mitf"uhrungskoeffizienten}. 
Fresnel konnte so eine wellentheoretische Erkl"arung des Nullresultats Aragos 
geben, ohne die Erkl"arung der stellaren Aberration zu gef"ahrden;
denn f"ur die Erdatmosph"are ist $n$ nahe eins und somit $f$ nicht me"sbar 
von Null verschieden. Sp"ater haben Fizeau~\cite{Fizeau:1851} im Jahre 
1851 und Michelson-Morley~\cite{MichelsonMorley:1886} im Jahre 1886 
die Beziehung (\ref{eq:FresnelKoeff}) experimentell best"atigt.
Schlie"slich konnte Hendrik Antoon Lorentz (1853-1928) im Jahre 1895 
im Rahmen seiner "`Elektronentheorie"' zeigen, da"s man den Ausdruck 
(\ref{eq:FresnelKoeff}) f"ur die anteilige Erh"ohung der 
Lichtgeschwindigkeit auch unter der Annahme eines absolut ruhenden 
"Athers erhalten kann~\cite{Lorentz:1895}. Die SRT wird schlie"slich 
eine von jeglicher "Athervorstellung unabh"angige Begr"undung von 
(\ref{eq:FresnelKoeff}) geben, die lediglich auf einer Modifikation 
der kinematischen Regel (\ref{eq:GeschAddGal}) beruht. Wir werden 
darauf noch zur"uckkommen.

\section{Die Maxwellsche Elektrodynamik}
\label{sec:Maxwell}
Die Lorentzsche "`Elektronentheorie"' ist eine Konkretisierung 
der Maxwellschen Elektrodynamik, der ersten umfassenden mathematischen 
Theorie elektrischer und magnetischer Ph"anomene, die der Schotte 
James Clerk Maxwell (1831-1879) in den 60er Jahren des 
19.\,Jahrhunderts formulierte und 1873 in seiner 
ber"uhmten "`Treatise on Electricity and Magnetism"' publizierte. 
Er hielt sich dabei eng an die Vorstellung des Chemikers und 
Experimentalphysikers Michael Faraday (1791-1867), der bei seiner 
Beschreibung elektrischer und magnetischer Wirkungen den noch heute 
beliebten Begriff der "`Kraftfeldlinien"' benutzte. Darunter verstand 
Faraday aber nicht nur eine abstrakte r"aumliche Verteilung von 
Kraftvektoren, sondern schrieb ihnen eine eigene physikalische 
Existenz zu, die unabh"angig vom lokalen Vorhandensein einer 
Test-Einheitsladung zur Messung der Kraft bestehen sollte. 
Er f"uhrte damit ein neues Realit"atskonzept in die Physik ein, 
das des \emph{elektrischen bzw. magnetischen Feldes}. 

Eine der sch"onsten Leistungen der Maxwellschen Theorie war die 
Voraussage elektromagnetischer Wellen, die sich im ladungs- und 
stromfreien (aber vom "Ather erf"ullt gedachten) Raum ausbreiten 
k"onnen. Die Ausbreitungsgeschwindigkeit wurde ebenfalls durch die 
Theorie vorhergesagt und ergab sich gleich der Lichtgeschwindigkeit. 
Damit setzte die Vorstellung ein, da"s Licht nichts anderes sei als 
eine elektromagnetische Welle und da"s die Gesetze der Optik, wie zum 
Beispiel die Brechungs- und Reflexionsgesetze, s"amtlich aus der 
Maxwell-Theorie folgen sollten, was sich dann auch im Verlauf des 
ausgehenden 19.\,Jahrhunderts gl"anzend best"atigte. 
Diese Tatsache, sowie die aufsehenerregenden Versuche, mit denen 
Heinrich Hertz (1857-1894) im Jahre 1888 an der Universit"at 
Karlsruhe direkt die Existenz elektromagnetischer Wellen nachwies, 
verhalfen Maxwells Theorie zum Durchbruch. Damit setzte sich auch 
das Konzept des lokalen Feldes als physikalische Zustandsbeschreibung
durch, wenn auch (noch) nicht in der modernen, emanzipierten 
Auffassung  als eigenst"andige Materieform, sondern stets gebunden 
an den substantiell konzipierten "Ather. Andernfalls h"atte man sich 
folgenden Fragen ausgesetzt gesehen, f"ur die es keine offensichtlichen 
Antworten gab: Was soll es bedeuten, eine Feldvariable einem abstrakten 
Raumpunkt ohne jedwede materielle Individualit"at zuzuordnen? 
Woran kann das Feld dann "`haften"'? Oder anders ausgedr"uckt: 
Wenn das Feld eine Zustandsgr"o"se sein soll, \emph{wessen} Zustand 
ist hier gemeint? 

\begin{figure}
\centering\epsfig{figure=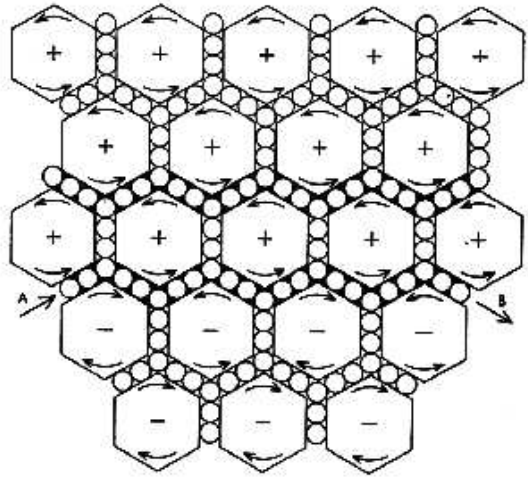, width=0.55\linewidth}
\label{fig:MaxwellAether}
\caption{Reproduktion einer Zeichnung Maxwells zur mechanischen 
Deutung des "Athers. Magnetfelder entstehen durch "`Molekularwirbel"', 
die gegenseitig durch die Anlagerung geladener Teilchen wie in einem 
Kugellager in Position gehalten werden. Ein Ladungstransport findet 
dann statt, wenn benachbarte Wirbel verschieden schnell rotieren. }
\end{figure}

Nachdem die Wesensgleichheit von Licht und elektromagnetischen 
Wellen einmal erkannt war, identifizierte man automatisch auch 
den das Licht tragenden "`"Ather"' mit dem die elektromagnetischen 
Kraftwirkungen vermittelnden Medium. "Uber die Natur dieses gemeinsamen 
"Athers war damit freilich noch nichts weiter ausgesagt, als da"s 
dessen Dynamik durch die Maxwell-Gleichungen sehr genau beschrieben 
wird. Die meisten glaubten, da"s sich auf fundamentalem Niveau 
diese Dynamik (und damit die Maxwell-Gleichungen) auf die 
Prinzipien der Mechanik w"urden zur"uckf"uhren lassen.\footnote{Es ist 
bemerkenswert, da"s auch Heinrich Hertz in seinem letzten, 
posthum herausgegebenen Werk "uber Mechanik (\cite{Hertz:GW}, Band\,III)
diesen Ansatz vertritt.} Dazu wurden verschiedene mechanische 
"Athermodelle vorgeschlagen, sogar von Maxwell selbst, wie das in 
Abbildung\,\ref{fig:MaxwellAether} von ihm selbst skizzierte. 

Somit blieb es auch nach Aufstellung der Maxwellschen Theorie 
bis zu Anfang des 20.\,Jahr\-hunderts bei einem dualistischen 
Materiebegriff. Dieser umfa"ste einerseits die im Raum lokal 
beweglichen "`ponderablen"' K"orper, andererseits den  
"Ather, der den ganzen Raum einschlie"slich des Inneren der 
K"orper durchdringt und Tr"ager elektromagnetischer Felder und damit  
auch der Lichtwellen ist. Auch das Gravitationsfeld 
war damals als in einem "`"Ather"' verankert zu denken 
(demselben?), doch war die Gravitationstheorie zu diesem 
Zeitpunkt noch sehr unterentwickelt, nicht vergleichbar etwa 
der Maxwellschen Theorie des Elektromagnetismus.

\section{Gilt das Relativit"atsprinzip in der Elektrodynamik\,?\\
Die "`Aarauer Frage"' und fr"uhe Skepsis}
\label{sec:ElekBewKoe}
Die Maxwellsche Theorie sagt f"ur die Geschwindigkeit 
elektromagnetischer Wellen einen bestimmten Wert voraus. 
Im "Atherbild ist dies nicht anders zu deuten als so, da"s dies 
die Geschwindigkeit relativ zum ruhend gedachten "Ather ist. 
Nur im Ruhesystem des "Athers sind also die Maxwellschen Gleichungen 
g"ultig.  Dies legt den Schlu"s nahe, da"s das Relativit"atsprinzip 
nicht auf die Elektrodynamik erweiterbar ist.  

Genau diese Schwierigkeit ahnte bereits der erst 16-j"ahrige Albert 
Einstein, damals Sch"uler der Aargauischen Kantonsschule. 
In einem R"uckblick aus dem Jahr seines Todes (1955) erinnerte 
sich Einstein an folgende Frage, die ihn als Sch"uler dieser 
Schule einst besch"aftigte und die Abraham Pais in seiner 
Einstein-Biographie~\cite{Pais:Einstein} die \emph{Aarauer Frage}
nannte:
\begin{quote}
"`W"ahrend dieses Jahres [zwischen Oktober 1895 und Fr"uhjahr 1896]
in Aarau kam mir die Frage: Wenn man einer Lichtwelle mit 
Lichtgeschwindigkeit nachl"auft, so w"urde man ein 
zeitunabh"angiges Wellenfeld vor sich haben. So etwas scheint 
es aber doch nicht zu geben! Dies war das erste kindliche 
Gedanken-Experiment, das mit der speziellen Relativit"atstheorie
zu tun hat."'
\end{quote}
Vergegenw"artigen wir uns dieses Argument: Laufen wir einer 
Lichtwelle, die sich im Ruhesystem des "Athers mit einer 
Geschwindigkeit $c$ ausbreitet, mit derselben Geschwindigkeit 
hinterher, so hat sie relativ zu uns eine verschwindende 
Ausbreitungsgeschwindigkeit. Hier wird offensichtlich das 
klassische Gesetz (\ref{eq:GeschAddGal}) f"ur die Komposition 
zweier Geschwindigkeiten benutzt. Dann haben wir also eine 
zeitunabh"angige (weil nicht fortschreitende), stehende Welle 
vor uns. Eine solche ist aber \emph{nicht} L"osung der Maxwellschen 
Gleichungen; also kann f"ur letztere das Relativit"atsprinzip auch 
nicht gelten. 

Erinnern wir uns an die Galileische Schilderung des Relativit"atsprinzips
in Abschnitt\,\ref{sec:RelPrinzTraeg}. Wie dort bereits betont, ist 
es f"ur seine G"ultigkeit wesentlich, da"s \emph{alle} physikalisch 
wirksamen Komponenten die gleichf"ormige Translationsbewegung 
mitmachen. Insbesondere mu"s die Luft im Inneren des Schiffes an dessen  
Bewegung vollst"andig teilnehmen, da sonst aufgrund des Luftzuges die 
Schmetterlinge eine anisotrope Geschwindigkeitsverteilung aufweisen 
w"urden, anhand derer eine Bewegung des Schiffes leicht auszumachen 
w"are. Die Rolle der Luft  "ubernimmt in der Elektrodynamik der "Ather. 
W"urde der "Ather alle Formen von Materie mehr oder weniger ungehindert 
"`durchwehen"' k"onnen, wie es die Fresnelsche Theorie 
zu fordern schien, dann g"abe es auch kein gegen"uber der 
"Atherbewegung v"ollig abgeschlossenes System und damit keine 
Verallgemeinerung des Relativit"atsprinzips auf die Elektrodynamik.
Es entsteht also die Frage, anhand welcher Wirkungen man den 
"Atherwind innerhalb der Materie direkt nachweisen kann.  

Bevor ich mich dieser Frage zuwende, m"ochte ich einige historische 
Bemerkungen machen, die belegen sollen, da"s die Besch"aftigung des 
jungen Einstein mit dieser Problematik weit "uber die "`Aarauer 
Frage"' hinausgingen. In einem Brief, den der 20-J"ahrige im September 
1899 an seine damalige Freundin und sp"atere erste Frau Mileva 
Mari\'c (1975-1948) schrieb, hei"st es (\cite{Einstein-CW}, 
Band\,1, Dokument\,54): 
\begin{quote}
"`In Aarau ist mir eine gute Idee gekommen zur Untersuchung,
welchen Einflu"s die Relativbewegung der K"orper gegen den 
Licht"ather auf die Fortpflanzungsgeschwindigkeit des Lichtes 
in durchsichtigen K"orpern hat. Auch ist mir eine Theorie in 
den Sinn gekommen "uber diese Sache, die mir gro"se 
Wahrscheinlichkeit zu besitzen scheint. Doch genug davon!"'
\end{quote}  
Genaueres "uber das von Einstein erdachte Experiment ist nicht 
bekannt, man kann aber vermuten, da"s es vom Grundgedanken her 
nicht un"ahnlich dem Fizeaus aus dem Jahre 1851 gewesen sein wird. 

Einstein berichtet Mileva "uber sein eifriges Studium der Schriften 
Helmholz' und Hertz'. Letzterer hatte 1890 in zwei viel 
beachteten Publikationen die Maxwellsche Theorie mathematisch 
ausgearbeitet~\cite{Hertz:1890a,Hertz:1890b}, auch f"ur den Fall 
bewegter K"orper. Dazu stellte er die Arbeitshypothese auf, da"s 
der "Ather im Inneren der K"orper an deren Bewegung vollst"andig 
teilnimmt. Wir werden darauf noch ausf"uhrlich zu sprechen kommen. 
Was uns hier interessiert ist 
Einsteins fr"uhe, allgemein skeptische Einsch"atzung der physikalischen 
Relevanz solcher Annahmen. Einen Monat vor dem eben zitierten Zeilen 
schrieb Einstein an Mileva (\cite{Einstein-CW}, Band\,1, Dokument\,52): 
\begin{quote}
"`Die Einf"uhrung des Namens "<"Ather"> in die elektrischen 
Theorien hat zur Vorstellung eines Mediums gef"uhrt, von dessen 
Bewegung man sprechen k"onne, ohne da"s man, wie ich glaube, mit 
dieser Aussage einen physikalischen Sinn verbinden kann."'
\end{quote} 
Genau dieser Glaube, konsequent umgesetzt, wird ihn 6 Jahre sp"ater 
zur SRT f"uhren.

\section{Experimente}
\label{sec:Experimente}
Zu der oben formulierten Frage, wie ein m"oglicher "Atherwind innerhalb 
bewegter Materie nachzuweisen sei, sind im Laufe der Entwicklung der 
Theorie des Lichtes und des Elektromagnetismus zahlreiche Versuche 
angestellt worden. Von diesen wollen wir hier stellvertretend (wie so oft) 
nur das Experiment von Michelson \& Morley 
referieren.\footnote{Weder nennt Einstein dieses noch andere 
Experimente namentlich in seiner 
Originalver"offentlichung~\cite{Einstein:SRT} zur SRT, 
so da"s mitunter kritisiert wird, dieses so zu pr"asentieren, als 
sei es der logische Ausgangspunkt der SRT gewesen. Einstein selbst 
hat sp"ater widerspr"uchliche Angaben dar"uber gemacht, ob er von 
diesem Experiment im Jahre 1905 "uberhaupt wu"ste~\cite{Shankland:1963}.
Sicher ist aber, da"s er die Lorentzsche Monographie~\cite{Lorentz:1895} 
aus dem Jahre 1895 gut kannte, in dem das Michelson-Morley-Experiment 
im Abschnitt~IV unter "`Versuche, deren Ergebnisse sich nicht ohne 
weiteres erkl"aren lassen"' als zweites Beispiel besprochen wird, 
so da"s man davon ausgehen kann, da"s Einstein zumindest die 
dortigen Ausf"uhrungen (bewu"st oder unbewu"st) gel"aufig waren.
(Das best"atigt auch ein Brief Einsteins an Robert Shankland vom 
Dezember 1952, der in \cite{Pais:Einstein} p.\,112 auszugsweise 
abgedruckt ist, sowie seine Gru"sadresse zu Michelsons Geburtstag
im Jahre 1952, deren Wortlaut in \cite{Shankland:1964} p.\, 35 
abgedruckt ist.) 
Wie auch immer, logischer Ausgangspunkt der SRT waren u.a., wie 
Einstein schreibt, "`die mi"slungenen Versuche, eine Bewegung der 
Erde relativ zum "<Lichtmedium"> zu konstatieren"', und zu diesen 
geh"ort an herausragender Stelle unstreitig das 
Michelson-Morley-Experiment. Eine umfassende und verst"andliche 
Diskussion historischer Experimente gibt \cite{Born:RT}. 
Die interessante Vorgeschichte des Michelson-Morley-Experiments 
wird in \cite{Shankland:1964} erz"ahlt.}
\begin{figure}[htb]
\noindent
\centering\epsfig{figure=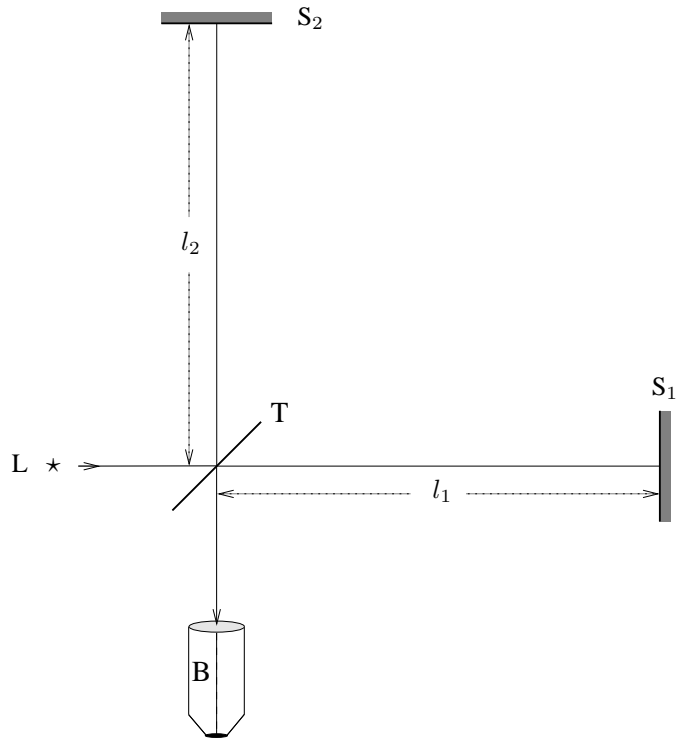, width=0.6\linewidth}
\put(-237,101){$\star$}
\put(-250,101){\small L}
\put(-142,270){\small $\mbox{S}_2$}
\put(-8,130){\small $\mbox{S}_1$}
\put(-186,185){\small$l_2$}
\put(-91,91){\small $l_1$}
\put(-152,120){T}
\put(-182,22){B}
\caption{\small Das Michelson-Morley-Experiment}
\label{fig:Michelson1}
\end{figure}
Dabei handelt es sich um ein sogenanntes Interferometer; siehe 
Abbildung\,\ref{fig:Michelson1}. 
Eine Quelle $L$ schickt einen Lichtstrahl auf einen Strahlteiler $T$, 
der die eine H"alfte des ankommenden Lichts durchl"a"st, so da"s es
"uber eine Wegstrecke $l_1$ zu einem Spiegel $S_1$ gelangt, von wo 
aus dieses zu $T$ zur"uckreflektiert wird und nach Reflexion an $T$ 
zum Teil zum Beobachter $B$ gelangt. Die andere H"alfte wird gleich 
von $T$ reflektiert, gelangt dann nach einer Wegstrecke $l_2$ zum 
Spiegel $S_2$, wird von dort zu $T$ zur"uckreflektiert und gelangt 
dann nach Passieren von $T$ teilweise zu $B$. Die zur Quelle $L$ 
zur"uckgelangenden Teilstrahlen interessieren nicht weiter. 
In $B$ befindet sich eine Vorrichtung, mit der das Interferenzmuster 
der dort "uberlagerten Teilstrahlen vermessen werden kann. 
Die ganze Anordnung ist mit allen Teilen fest auf einer Steinplatte 
montiert, die ihrerseits in einer Wanne mit Quecksilber schwimmt, so 
da"s sie im Ganzen m"oglichst ersch"utterungsfrei gedreht werden kann. 
Da die ganze Apparatur auf der Erdoberfl"ache ruht, f"uhrt sie 
relativ zur Sonne eine Bewegung aus, die sich aus der t"aglichen 
Eigenrotation der Erde und ihrer j"ahrlichen Bahnbewegung um die 
Sonne zusammensetzt. 

Messungen der Lichtausbreitung in str"omenden Fl"ussigkeiten, die 
Armand Fizeau (1819-1896) im Jahre 1951 anstellte, hatten die bereits 
erw"ahnte Fresnelsche Theorie von 1818 best"atigt. Gem"a"s dieser sollte 
die Phasengeschwindigkeit des Lichts durch das str"omenden Medium 
(Brechungsindex $n$) einen Zuwachs $\Delta v$ in Str"omungsrichtung 
nur um den Bruchteil $f$ (siehe (\ref{eq:FresnelKoeff})) der 
Str"omungsgeschwindigkeit erhalten:\footnote{Interessanterweise 
wurde zur damaligen Zeit gerade diese Tatsache der nur \emph{teilweisen} 
Mitf"uhrung als schlagender Beweis f"ur die Existenz eines "Athers 
angesehen. Das Argument verlief etwa so: G"abe es keinen "Ather und 
w"are also nur das materielle Umgebungsmedium f"ur die Ausbreitung 
des Lichtes ma"sgebend, so m"u"ste wegen des (f"ur unzweifelhaft 
gehaltenen) Kompositionsgesetzes (\ref{eq:GeschAddGal}) f"ur 
Geschwindigkeiten eine volle Mitf"uhrung eintreten. In der SRT ist 
(\ref{eq:Fizeau}) nun eine direkte Folge des ge"anderten 
Kompositionsgesetzes f"ur Geschwindigkeiten; siehe~(\ref{eq:FresnelSRT}).}   
\begin{equation}
\label{eq:Fizeau}
\Delta v=v\cdot f=v\cdot\bigl(1-n^{-2}\bigr)\,.
\end{equation}

Interpretiert man dies als die Mitf"uhrungsgeschwindigkeit des "Athers,
so sollte wegen der N"ahe des Brechungsindex' der Luft zum Wert 1 
keine merkliche Mitf"uhrung des "Athers durch die sich mit der Erde 
bewegende Erdatmosph"are zu erwarten sein. Also sollte im 
Laufe eines Jahres die Geschwindigkeit der Michelson-Morley-Apparatur 
relativ zum "Ather einmal mindestens so gro"s werden wie die 
Relativgeschwindigkeit der Erde zur Sonne, d.h. 30 Kilometer 
pro Sekunde. Im Michelson-Morley-Experiment wird nun "uberpr"uft, ob 
die Differenz zwischen der f"ur Hin- und R"ucklauf zusammen ben"otigten 
Lichtlaufzeit des einen Arms zu der des anderen Arms von der Orientierung der 
Apparatur abh"angt. Nach der "Athertheorie ist eine solche Abh"angigkeit 
n"amlich zu erwarten, wie die folgende Diskussion zeigt.   

Wir verfolgen den Vorgang der Lichtausbreitung vom Ruhesystem des "Athers 
aus. Aus Gr"unden der Einfachheit nehmen wir an, da"s sich die Apparatur 
momentan mit dem Geschwindigkeitsbetrag $v$ in Richtung des ersten Strahls, 
also in Richtung $TS_1$, relativ zum "Ather bewegt. Von $T$ nach $S_1$ 
l"auft der Lichtstrahl dem "`Atherwind entgegen und hat daher relativ zur 
Apparatur die Geschwindigkeit $c-v$ (hier wenden wir (\ref{eq:GeschAddGal}) 
an). Auf dem Weg zur"uck von $S_1$ nach $T$ hat er den "Atherwind 
"`im R"ucken"' und hat dementsprechend die Relativgeschwindigkeit 
$c+v$ zur Apparatur. Zusammen ergibt sich eine Laufzeit von $T$ nach 
$S_1$ und zur"uck von ($\gamma$ ist wie in (\ref{eq:LorentzTrans3}) 
definiert): 
\begin{equation}
\label{eq:MM1}
T_1=\frac{l_1}{c-v}+\frac{l_1}{c+v}=\frac{2l_1}{c}\,\gamma^2\,.  
\end{equation}

\begin{figure}[htb]
\noindent
\centering\epsfig{figure=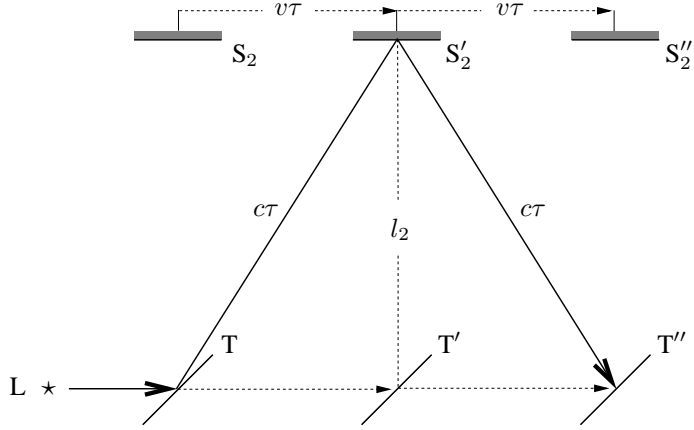, width=0.6\linewidth}
\put(-63,156){\small $v\tau$}
\put(-147,156){\small $v\tau$}
\put(-54,80){\small $c\tau$}
\put(-155,80){\small $c\tau$}
\put(-235,12){$\star$}
\put(-247,12){\small L}
\put(1,138){\small $\mbox{S}''_2$}
\put(-82,138){\small $\mbox{S}'_2$}
\put(-163,138){\small $\mbox{S}_2$}
\put(-2,28){\small $\mbox{T}''$}
\put(-85,28){\small $\mbox{T}'$}
\put(-167,28){\small $\mbox{T}$}
\put(-103,72){\small $l_2$}
\caption{\small Transversaler Strahlengang im Michelson-Morley-Experiment
vom Ruhesystem des "Athers aus gesehen.}
\label{fig:Michelson2}
\end{figure}

Zur Beurteilung der Lichtlaufzeit von $T$ nach $S_2$ und zur"uck betrachten 
wir Abbildung\,\ref{fig:Michelson2}.
Zu beachten ist, da"s sich das vertikal angeordnete Paar von Strahlteiler $T$
und zweitem Spiegel $S_2$ w"ahrend des Lichtlaufes selbst in horizontaler 
Richtung weiterbewegt. In Abbildung\,\ref{fig:Michelson2} bezeichnet $T,S_2$ 
dieses Paar zu dem Zeitpunkt, in dem der Lichtstrahl (genauer: eine 
bestimmte Phase) zum ersten Male $T$ trifft, $T',S'_2$ zum Zeitpunkt, 
in dem der Lichtstrahl (dieselbe Phase) $S_2$ erreicht und schlie"slich 
$T'',S''_2$ zu dem Zeitpunkt, in dem der Lichtstrahl (dieselbe Phase) 
zu $T$ zur"uckkehrt. Wir bezeichnen ferner mit $\tau$ die vom Licht 
ben"otigte Zeitspanne, um von $T$ zum Spiegel zu gelangen. 
Dieser hat sich in dieser Zeitspanne von $S_2$ nach $S'_2$ um die Strecke 
$v\tau$ weiterbewegt. Ebenso hat sich der Strahlteiler in der Zeitspanne 
$\tau$, die das Licht auf dem Weg zur"uck vom Spiegel $S_2'$ zum 
Strahlteiler ben"otigt, um die Strecke $v\tau$ von $T'$ nach $T''$ 
weiterbewegt. Wegen der Rechtwinkligkeit der Dreiecke $TT'S'_2$ bzw. 
$S'_2T'T''$ und der angegebenen Streckenl"angen ist nach dem Satz des 
Pythagoras $c^2\tau^2$ gleich der Summe $l_2^2+v^2\tau^2$, was 
leicht nach $\tau$ aufgel"ost werden kann. Daraus ergibt sich sofort 
die Laufzeit in der Vertikalkomponente:   
\begin{equation}
\label{eq:MM3}
T_2=2\tau= \frac{2l_2}{c}\,\gamma\,.  
\end{equation}
Man beachte, da"s sich dies von (\ref{eq:MM1}) dadurch 
unterscheidet, da"s $\gamma$ hier linear, dort aber quadratisch 
eingeht. 

Die selbst bei gleichen Arml"angen ($l_1=l_2$) unterschiedlichen 
Laufzeiten bedeuten, da"s eine bestimmte Phase der Lichtwelle, die 
zu einem festen Zeitpunkt von der Quelle $L$ emittiert wird, zu 
unterschiedlichen Zeiten beim Beobachter $B$ eintrifft, je nachdem, 
ob sie den Weg "uber $S_1$ oder "uber $S_2$ genommen hat. Betreibt 
man das Experiment mit monochromatischem Licht der Frequenz $\nu$, 
so hat nach Vereinigung bei $B$ das "uber den Spiegel $S_1$ laufende 
Licht die feste Anzahl 
\begin{equation}
\label{eq:MM4}
N=\nu(T_1-T_2)   
\end{equation}
von Phasen dem "uber $S_2$ laufenden Licht voraus, was sich in 
einem festen Interferenzmuster bei $B$ "au"sert. Dreht man nun die 
Anordnung im Uhrzeigersinn um 90~Grad, so zeigt $TS_2$ in die 
Bewegungsrichtung relativ zum "Ather und $TS_1$ senkrecht dazu. 
Durch einfache Wiederholung der obigen Argumentation ist klar, 
da"s die neuen Lichtlaufzeiten $T_1',T_2'$ "uber $S_1$ bzw. $S_2$
im gedrehten Zustand gegeben sind durch 
\begin{equation}
\label{eq:MM5}
T'_1=\frac{2l_1}{c}\,\gamma \quad\mbox{und}\quad
T'_2=\frac{2l_2}{c}\,\gamma^2\,.  
\end{equation}
Die Anzahl $N'$ der Phasen, die das "uber den Spiegel $S_1$ 
laufende Licht nach Vereinigung bei $B$ dem "uber $S_2$ laufenden 
voraus hat, ist nun gegeben durch
\begin{equation}
\label{eq:MM6}
N'=\nu(T'_1-T'_2)\,.
\end{equation}
Die Differenz $N-N'$ entspricht daher gerade der Anzahl der 
Streifen, um die sich das Interferenzmuster in $B$ w"ahrend des 
Vorgangs der Drehung um 90 Grad verschiebt. Dr"uckt man die 
Lichtfrequenz $\nu$ noch durch die Lichtwellenl"ange $\lambda=c/\nu$
aus, so erh"alt man f"ur diese Anzahl 
\begin{equation}
\label{eq:MM7}
\Delta N=N-N'=2\ \frac{l_1+l_2}{\lambda}\cdot \gamma(\gamma-1)
          \approx\frac{l_1+l_2}{\lambda}\cdot\frac{v^2}{c^2}\,,
\end{equation}
wobei der zweite Ausdruck der rechten Seite in sehr guter 
Approximation f"ur solche Geschwindigkeiten gilt, die nicht allzu 
nahe der Lichtgeschwindigkeit liegen (wir haben vierte und h"ohere 
Potenzen in $v/c$ vernachl"assigt). Man beachte, 
da"s $v/c$ in (\ref{eq:MM7}) quadratisch eingeht, also der Effekt 
selbst f"ur kleine $v/c$ stark unterdr"uckt ist, verglichen etwa mit 
Effekten linearer Ordnung wie der Aberration (vgl. etwa \cite{Giulini:SRT}). 
Setzt man entsprechend der Bahngeschwindigkeit der Erde  $v/c=10^{-4}$, 
so ist $v^2/c^2=10^{-8}$\,!  
Michelson-Morley verwendeten gleiche Arml"angen von effektiv (sie wurden
in diesem Experiment vom Licht tats"achlich mehrfach durchlaufen, bevor sie 
zur Interferenz gebracht wurden) 11 Metern und Licht der Wellenl"ange von 
5900\,{\AA}ngstr"om ($5,9\cdot 10^{-7}m$) (was einem gelben Farbton 
entspricht), so da"s sie erwarteten, eine Verschiebung von $\Delta N=0{,}37$ 
zu sehen. Dabei war die Empfindlichkeit ihrer Apparatur so hoch, da"s sie 
noch Streifenverschiebungen von einem Hundertstel h"atten messen k"onnen, 
also fast einem Vierzigstel des zu erwartenden Effekts.

Das "uberraschende Ergebnis ihrer Messungen war jedoch, da"s sie im Rahmen 
dieser Genauigkeit \emph{keine} Verschiebungen ma"sen. F"ur die bisher 
beschriebene "Athertheorie bedeutete dies, da"s die Relativgeschwindigkeit
der Erde zum "Ather wesentlich kleiner sein mu"ste als die 
Bahngeschwindigkeit der Erde um die Sonne. Doch selbst wenn zuf"allig die 
Erde zum Zeitpunkt des Experiments relativ zum "Ather ruht, so sollte nach 
einem halben Jahr ihre Relativgeschwindigkeit zum "Ather sogar 60 Kilometer 
pro Sekunde betragen, da sie sich dann auf der diametral gegen"uberliegenden 
Seite ihrer Bahn um die Sonne befindet, wo ihre Bewegungsrichtung genau 
gegenl"aufig ist. Deshalb ist das Experiment auch zu verschiedenen 
Jahreszeiten wiederholt worden, doch das Ergebnis war auch dann stets 
negativ. Die Resultate modernster Experimente sind z.B. 
in \cite{Giulini:SRT} quantitativ wiedergegeben. 

Auch wurde gemutma"st, da"s vielleicht doch eine Mitf"uhrung des "Athers 
stattfindet, bedingt einerseits durch die tieferen und damit dichteren 
Schichten der Erdatmosph"are (wogegen allerdings die Resultate Fizeaus
sprechen w"urden), vor allem aber durch die festen, undurchsichtigen  
Geb"audew"ande, die das Experiment umgaben.\footnote{Das Experiment wurde 
absichtlich in einem Keller mit starken W"anden ausgef"uhrt, um es so 
von "au"seren Einfl"ussen fernzuhalten.} So hat Dayton Miller (1866-1941), 
ein fr"uherer 
Mitarbeiter Morleys, sogar noch 1921 das Michelson-Morley-Experiment in 
einer d"unnwandigen Baracke auf dem Gipfel des Mount Wilson wiederholt, 
um einen m"oglichst "`ungebremsten"' Durchzug des "Atherwindes zu 
gew"ahrleisten. Und tats"achlich schien seine erste Messung ein positives 
Resultat zu liefern, entgegen den Voraussagen der damals schon 16 Jahre 
bestehenden Speziellen Relativit"atstheorie, was Einstein mit seinem 
seither ber"uhmten Spruch kommentierte: "`Raffiniert ist der Herrgott, 
aber boshaft ist er nicht"'.\footnote{Die ganze Geschichte um das 
scheinbar positive Resultat Millers erregte damals einiges Aufsehen, so 
da"s Max Born auf seiner Amerikareise 1925/26 Miller und seine 
Versuchsanordnung in Augenschein nahm. Nach Angaben von Frau Born 
war ihr Mann "`entsetzt von der lodderigen Versuchsanordnung"'. 
Er selbst schrieb: "`Ich fand, da"s alles ganz wackelig und unzuverl"assig 
war; die kleinste Handbewegung oder ein Husten machte die 
Inteferenzstreifen so unruhig, da"s von Ablesen keine Rede sein konnte. 
Danach glaubte ich "uberhaupt nicht an Millers Ergebnisse. Ich kannte 
ja von meinem Chicagoer Aufenthalt im Jahre 1912 die Zuverl"assigkeit 
von Michelsons eigenen Apparaten und die Genauigkeit seiner Messungen."'}
In der Tat konnten Miller und seine Nachfolger das scheinbar positive 
Resultat nicht mehr reproduzieren, so da"s  man von einer Fehlmessung 
ausgehen mu"s, zumal die von ihm erreichte Genauigkeit wegen der 
beabsichtigten schlechten Abschirmung, die das Experiment allerlei 
Umwelteinfl"ussen aussetzte, nur etwa ein Drittel der des urspr"unglichen 
Experiments von 1887 war. 

Zusammenfassend entstand also die etwas paradoxe Situation, da"s man 
einerseits gezwungen war, einen durch Materie nur partiell mitgef"uhrten 
"Ather anzunehmen, andererseits aber keine experimentelle Methode 
in der Lage zu sein schien, eine Relativbewegung zum "Ather 
direkt nachzuweisen.

\section{Ein scheinbar verwegener Erkl"arungsversuch}
\label{sec:Erklaerung}
\begin{figure}[htb]
\begin{minipage}[c]{0.2\linewidth}
\centering\epsfig{figure=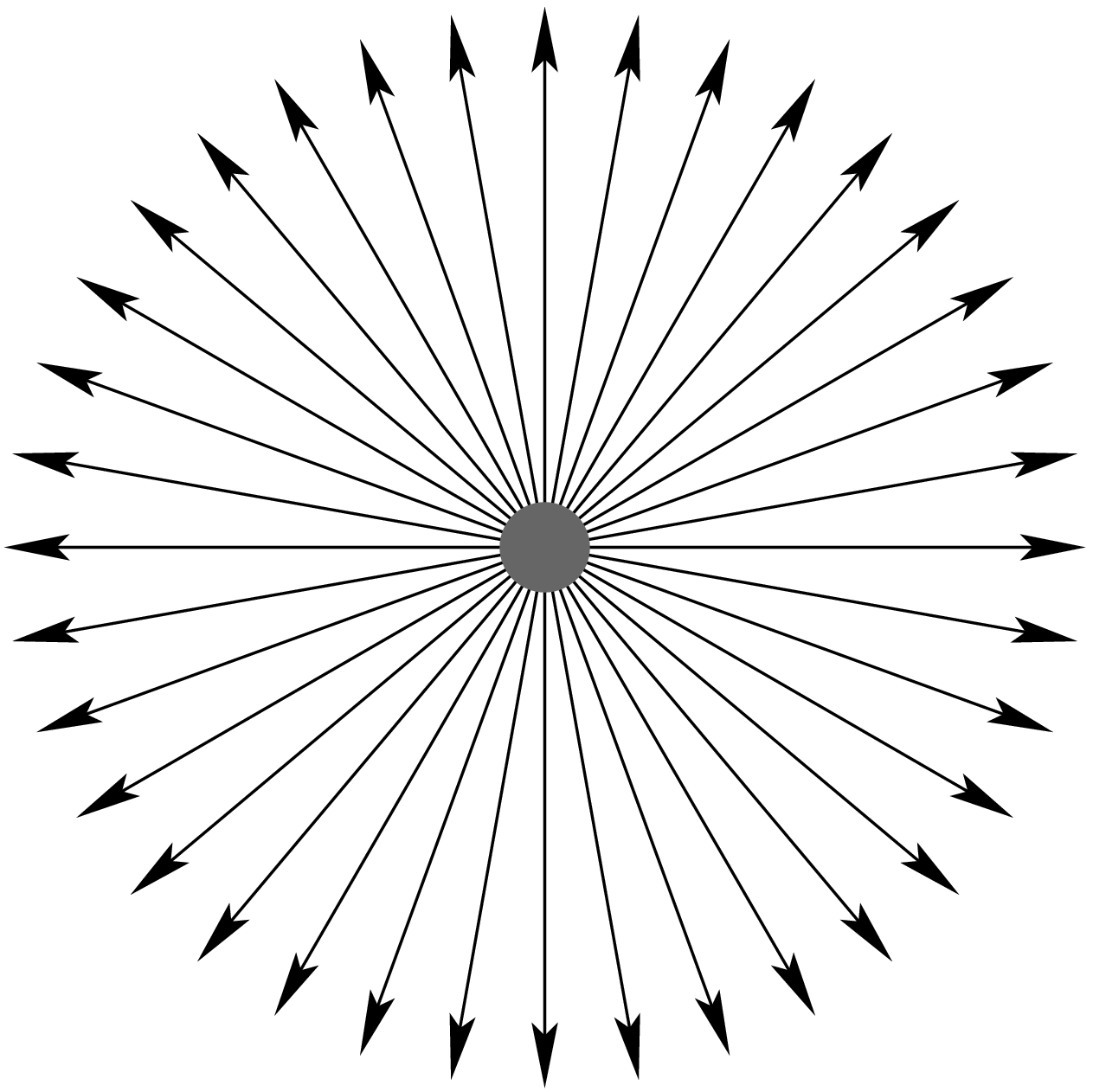, width=1.0\linewidth}
\end{minipage}
\hfill
\begin{minipage}[c]{0.2\linewidth}
\centering\epsfig{figure=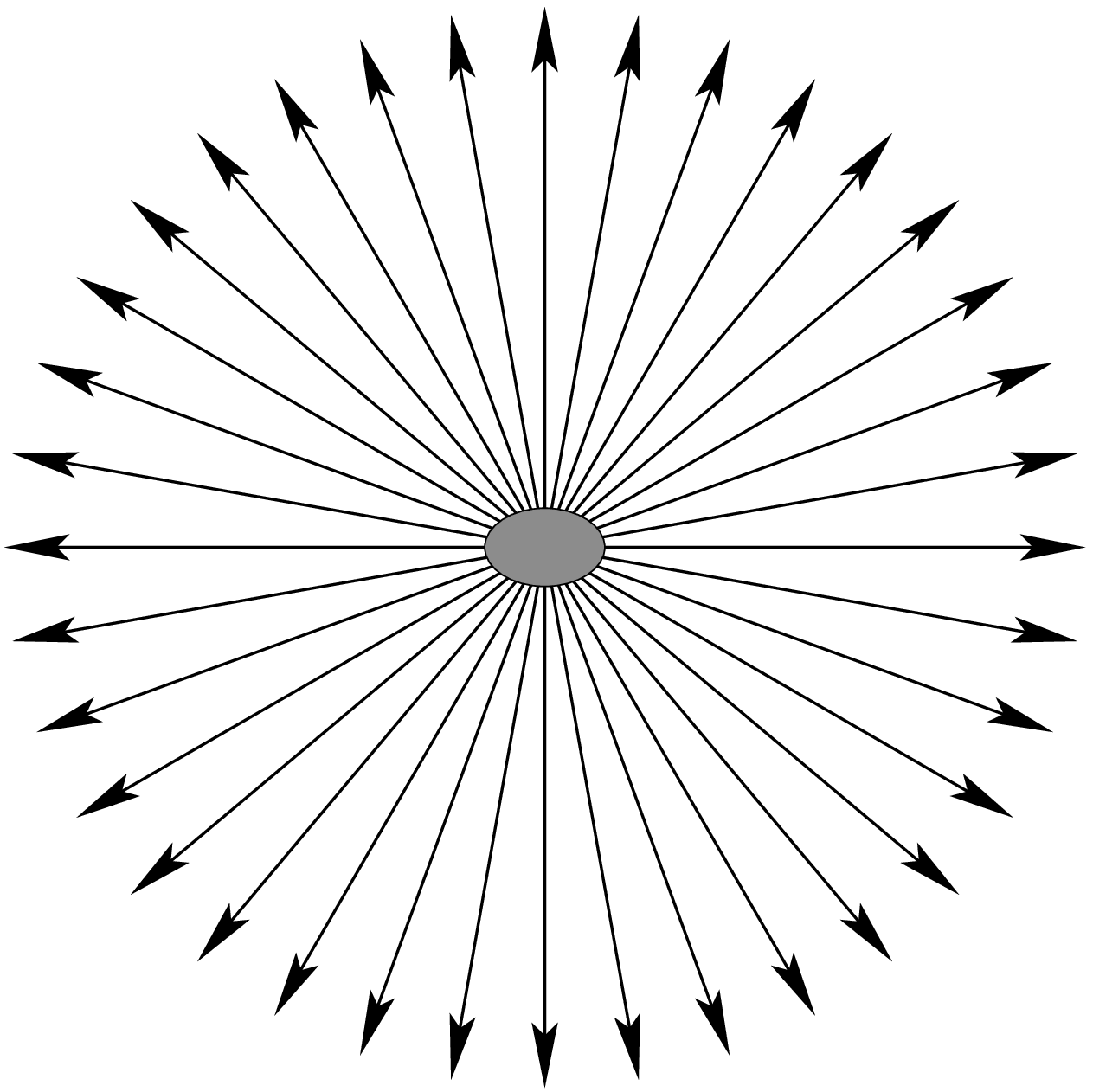, width=0.75\linewidth, 
                                      height=1.15\linewidth}
\end{minipage}
\hfill
\begin{minipage}[c]{0.2\linewidth}
\centering\epsfig{figure=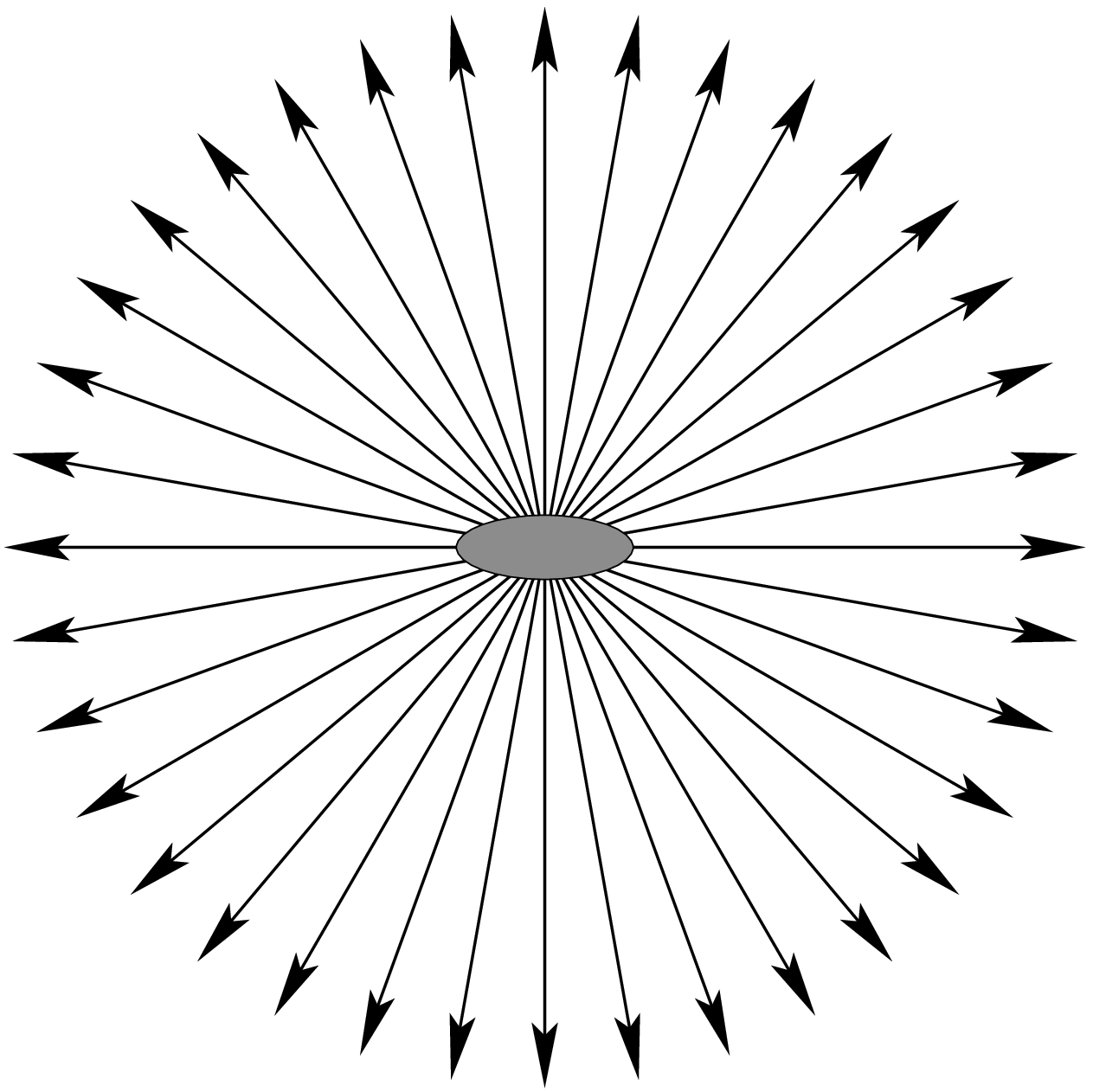, width=0.51\linewidth, 
                                      height=1.4\linewidth}
\end{minipage}
\hfill
\begin{minipage}[c]{0.2\linewidth}
\centering\epsfig{figure=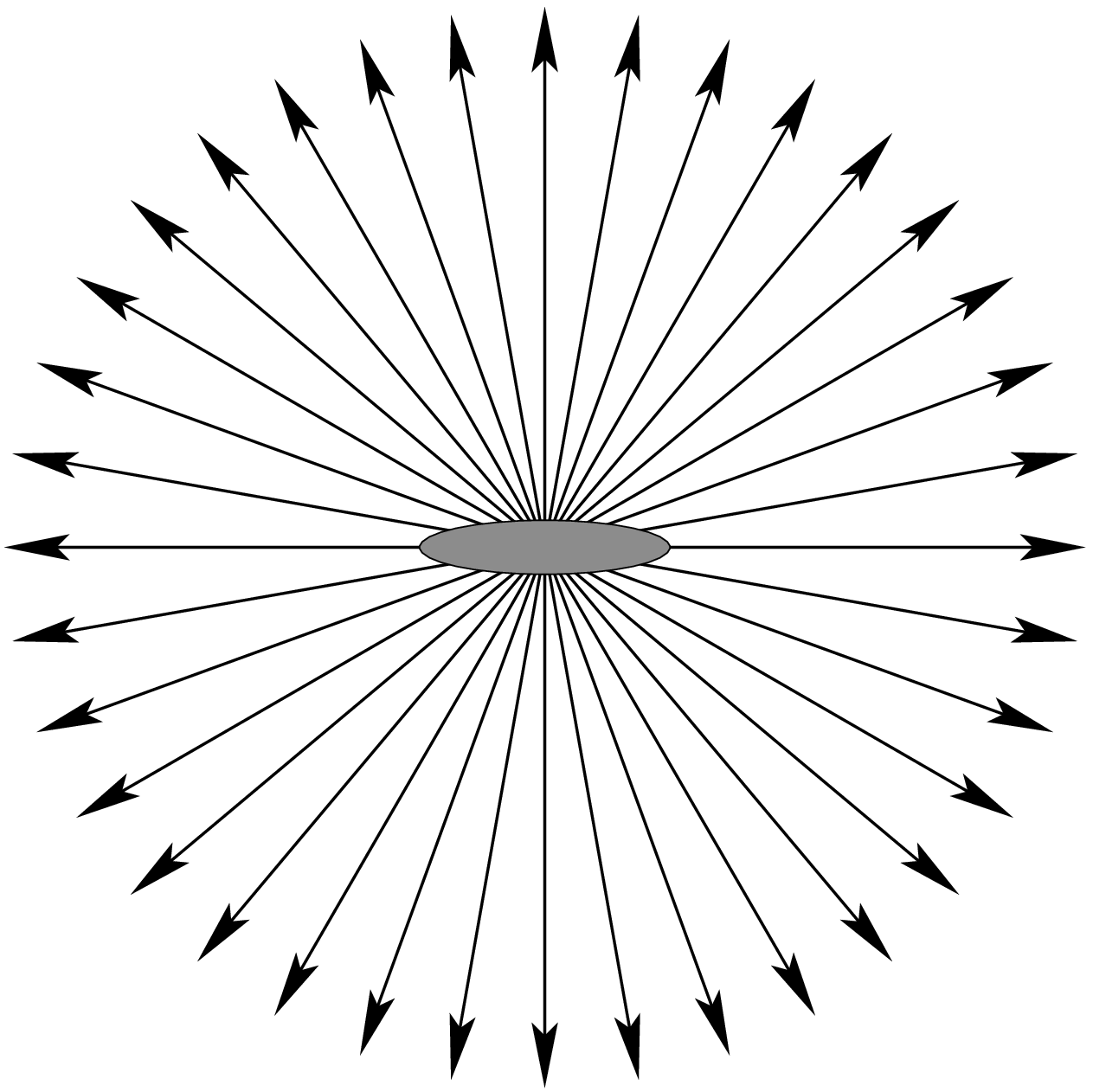, width=0.36\linewidth, 
                                      height=1.66\linewidth}
\end{minipage}
\caption{Coulombfeld einer Punktladung im Ruhezustand und bei 50, 70 
und 80 Prozent der Lichtgeschwindigkeit.}
\label{fig:Coulombfeld}
\end{figure}

Kaum zwei Jahre nach dem Michelson-Morley-Experiment erschien in der 
amerikanischen Fachzeitschrift "`Science"' eine halbseitige 
Note~\cite{FitzGerald:1889} des irischen Physikers George Francis 
FitzGerald (1851-1901) mit einem theoretischen Deutungsversuch des 
negativen Ausgangs. Darin stellte FitzGerald die zun"achst wild spekulativ 
anmutende Hypothese auf, da"s Ma"sverh"altnisse eines gegen"uber 
dem "Athersystem bewegten starren K"orpers als Folge dieser 
Relativbewegung eine \emph{universelle} -- d.h. vom Material und 
dessen physikalisch-chemischen Eigenschaften 
unabh"angige --, nur von der Relativgeschwindigkeit abh"angige 
Ver"anderung erleiden. FitzGeralds Note blieb weitgehend 
unbeachtet, bis im Jahre 1892 der bekanntere Lorentz, wohl 
unabh"angig von FitzGerald, mit einer fast identischen Idee 
aufwartete~\cite{Lorentz:1892}. Tats"achlich war eine solche 
Hypothese gar nicht so abwegig, wenn man sich auf einen atomistischen 
Standpunkt stellte und annahm, da"s die Konstitution eines jeden 
festen K"orpers ausschlie"slich durch die elektrostatischen 
Bindungskr"afte elementarer Kraftzentren (Atome oder Molek"ule) 
bestimmt ist. Es war n"amlich durch die Arbeiten von Oliver 
Heaviside (1850-1925) seit 1888 bekannt, da"s nach der 
Maxwellschen Theorie das elektrische Feld einer bewegten 
Punktladung gegen"uber dem einer ruhenden Punktladung in 
Bewegungsrichtung gestaucht ist, wie in Abbildung\,\ref{fig:Coulombfeld}
gezeigt. Dabei nahm man an, da"s die Maxwell-Gleichungen sich 
ausschlie"slich auf das Ruhesystem des "Athers beziehen, so da"s 
"`bewegt"' und "`ruhend"' relativ zum "Ather zu verstehen sind. 
Demnach k"onnte man mutma"sen, da"s sich auch ein starrer K"orpers 
in Bewegungsrichtung zusammenstaucht. Nat"urlich war es einstweilen 
unbekannt, ob tats"achlich alle im Inneren eines festen K"orper 
wirkenden Kr"afte elektromagnetischer Natur sind.

Um zu demonstrieren, wie diese Hypothese den Ausgang des 
Michelson-Morley-Experiments erkl"aren kann, denken wir uns die 
Anordnung der Abbildung\,\ref{fig:Michelson1} relativ zum "Ather 
zun"achst in Ruhe. Die Arml"angen seien in diesem Zustand mit 
$l_1^0$ und $l_2^0$ bezeichnet. Bei Bewegung relativ zum 
"Athersystem nehme man an, da"s alle L"angen in Bewegungsrichtung 
mit dem Faktor $A$, alle senkrecht dazu mit dem Faktor $B$ skaliert 
werden, wobei $A$ und $B$ vom Betrag der Geschwindigkeit abh"angen. 
Konkret bedeutet dies, da"s das Vermessen eines Objektes 
\emph{mit relativ zum "Athersystem ruhenden Ma"sst"aben} dieses Skalierungsverhalten 
zeigt. W"urde man hingegen das Objekt mit mitbewegten Ma"sst"aben 
vermessen, so w"urde man keine "Anderungen seiner Ma"se feststellen, 
da wegen der Universalit"at auch die Ma"sst"abe in gleicher Weise von 
diesem Effekt betroffen w"aren. Wie bei der obigen Diskussion des 
Michelson-Morley-Experiments werden auch im folgenden alle Angaben auf 
das "Athersystem bezogen. Setzt man nun die Anordnung relativ zum "Ather 
mit dem Geschwindigkeitsbetrag $v$ in Bewegung und bezeichnet man mit 
$l_1$ und $l_2$ die L"angen der Arme im Falle der Bewegung in Richtung 
des ersten Arms und mit $l_1'$ und $l_2'$ nach Drehung der Apparatur, 
so gilt zufolge der Hypothese
\begin{equation}
\label{eq:MM8}
A=\frac{l_1}{l_1^0}=\frac{l'_2}{l_2^0}
\quad\mbox{und}\quad
B=\frac{l_2}{l_2^0}=\frac{l'_1}{l_1^0}\,. 
\end{equation}
Benutzt man dies, um in obigen Ausdr"ucken (\ref{eq:MM1}) f"ur $T_1$ 
und (\ref{eq:MM3}) f"ur $T_2$ die Gr"o"sen $l_1$ bzw. $l_2$ durch 
$Al_1^0$ bzw. $Bl_2^0$ zu ersetzen und verf"ahrt man in derselben 
Weise in den Ausdr"ucken (\ref{eq:MM5}) f"ur $T'_1$ und $T'_2$, wobei 
dort wegen der jetzt ge"anderten Bezeichnungsweise zun"achst $l_1$
und $l_2$ als $l'_1$ und $l'_2$ geschrieben werden m"ussen, so erh"alt 
man anstatt (\ref{eq:MM7}) f"ur die Anzahl der verschobenen Streifen 
den Ausdruck
\begin{equation}
\label{eq:MM9}
\Delta N=2\ \frac{l_1^0+l_2^0}{\lambda}\ \gamma(\gamma A-B)\,.
\end{equation}
Der negative Ausgang des Michelson-Morley-Experiments, d.h. $\Delta N=0$, 
w"are also durch die Hypothese erkl"arbar, da"s der Skalierungsfaktor 
$A$ f"ur L"angen in Bewegungsrichtung das $1/\gamma$--Fache des 
Skalierungsfaktors $B$ f"ur L"angen senkrecht dazu ist, denn dann 
verschwindet der letzte Klammerausdruck auf der rechten Seite von 
(\ref{eq:MM9}). Insbesondere ist es hinreichend -- aber zur Erkl"arung 
des Michelson-Morley-Experiments nicht notwendig --, wenn man 
$A=1/\gamma$ und $B=1$ setzt, so da"s transversal zur Bewegungsrichtung 
keine L"angen"anderung stattfindet, in Bewegungsrichtung jedoch eine 
Verk"urzung (Kontraktion) mit dem Faktor $1/\gamma$. 
Wir betonen aber nochmals, da"s das Michelson-Morley-Experiment 
zur Festlegung der Werte von $A$ und $B$ nicht ausreicht. Dies gelingt 
erst durch Hinzunahme zweier weiterer Experimente 
(siehe z.B. \cite{Giulini:SRT}).

\section{Lorentz und Poincar\'e}
\label{sec:LorentzPoincare}
\subsection{Lorentz}
Die theoretische Weiterentwicklung dieser Idee wurde dann erneut 
durch Lorentz und Henri Poincar\'e (1854-1912) vorangetrieben. 
Dazu konkretisierte Lorentz die Maxwellsche Elektrodynamik in 
zweierlei Hinsicht: Erstens nahm er an, da"s der "Ather absolut 
unbeweglich sei, also keinerlei Mitf"uhrung innerhalb von Materie
stattfindet. Zweitens spezifizierte er -- was uns heute 
selbstverst"andlich erscheint, damals aber "uberhaupt nicht evident 
war -- , da"s es nur eine Art von elektrischen Str"omen gibt, 
n"amlich Konvektionsstr"ome elektrisch geladener Teilchen (Elektronen). 
Diese Teilchen sind zwar klein, aber endlich ausgedehnt zu denken 
und k"onnen sich frei im "Ather bewegen, der sie dabei reibungsfrei 
und unver"andert durchstr"omt.  Diese Theorie Lorentz' war weithin 
unter dem Namen "`Elektronentheorie"' bekannt. 

Wie schon erw"ahnt, gelang es Lorentz, in diesem Rahmen den 
Fresnelschen Mitf"uhrungskoeffizienten (\ref{eq:FresnelKoeff})
f"ur die Lichtgeschwindigkeit in bewegter Materie abzuleiten, 
was f"ur sich genommen bereits als gro"ser Erfolg gewertet wurde. 
Aber Lorentz ging viel weiter, indem er auf der Basis seiner 
Elektronentheorie argumentierte, da"s es "uberhaupt 
keinen elektromagnetischen Effekt linearer Ordnung in $v/c$ gibt, 
der die Bewegung relativ zum "Ather anzeigt. Um dies zu zeigen, 
mu"ste er jedoch annehmen, da"s die Orts- und Zeitkoordinaten 
des bewegten Systems (hier mit einem Strich versehen) mit denen 
des im "Ather ruhenden Koordinatensystems (ohne Strich) wie folgt 
verbunden sind:\footnote{Formal zeigte Lorentz Folgendes:
Es gibt eine lineare Transformation zwischen den auf das 
unbewegte System $K$ mit Koordinaten $(t,\vec x)$ bezogenen 
elektromagnetischen Feldern $F(t,\vec x)$ und den auf das bewegte 
System $K'$ mit Koordinaten $(t',\vec x')$ bezogenen Feldern 
$F'(t',\vec x')$, so da"s $F'(t',\vec x')$ die dynamischen 
Grundgleichungen im System $K'$ genau dann erf"ullt, wenn 
$F(t,\vec x)$ die dynamischen Grundgleichungen im System $K$ 
erf"ullt. Man sagt: die Gleichungen sind \emph{invariant} unter der 
Transformation $(t,\vec x)\mapsto (t',\vec x')$.} 
\begin{subequations}
\label{eq:LorentzTransApp}
\begin{alignat}{2}
\label{eq:LorentzTransApp1}
& \vec x'&&\,=\,\vec x-\vec vt\,,\\
\label{eq:LorentzTransApp2}
& t'&&\,=\,t-\vec v\cdot\vec x/c^2\,.
\end{alignat}
\end{subequations} 
Das sind gerade die Lorentz-Transformationen (\ref{eq:LorentzTrans}),
wenn man dort $\gamma=1$ setzt, also quadratische und h"ohere Potenzen
in $v/c$ vernachl"assigt.  F"ur Lorentz hingegen, der noch ganz auf 
dem Boden der Galilei-Transformationen stand,  war das Ungew"ohnliche 
daran der Unterschied zwischen (\ref{eq:GalileiTrans2}) und 
(\ref{eq:LorentzTransApp2}), also das nichttriviale 
Transformationsverhalten der Zeit. Dazu bemerkt er nur lapidar 
(\cite{Lorentz:1895}, p.\,50): 
\begin{quote}
"`Es sei hier noch die Bemerkung vorausgeschickt, da"s die Variable 
$t'$ als Zeit betrachtet werden kann, gerechnet von einem von der 
Lage des betreffenden Punktes abh"angigen Augenblick an. Man kann 
daher diese Variable die \emph{Ortszeit} dieses Punktes, im Gegensatz 
zu der \emph{allgemeinen Zeit} $t$, nennen."'
\end{quote}
Mit "`kann als Zeit betrachtet werden"' meint Lorentz wohl, da"s 
$t$ und $t'$ f"ur festes $\vec x$ gleich schnell laufen, 
d.h. $dt'/dt =1$. Lediglich der Nullpunkt der Zeitskala $t'$ 
mu"s in einer von $\vec x$ abh"angigen Weise gew"ahlt werden, 
was aber einer physikalisch unbedeutenden Konvention gleichkommt. 

Damit war die Nicht-Existenz sogenannter "`Effekte erster Ordnung"' 
gezeigt, aber nicht die quadratischer Ordnung, wie etwa im 
Michelson-Morley-Experiment (vgl. \ref{eq:MM7}). 
Diese konnten jedoch dadurch erledigt werden, da"s man der 
Elektronentheorie die oben erw"ahnte Kontraktionshypothese 
zus"atzlich aufpropfte. Um letztere in der Elektronentheorie zu 
verankern, nahm Lorentz an, da"s die Kontraktionshypothese f"ur die 
elementaren Ladungstr"ager (Elektronen) g"ultig sei. 
So kam Lorentz zu seiner Vorstellung des "`deformierbaren 
Elektrons"'. In der vielleicht wichtigsten Lorentzschen 
Vorl"auferarbeit zur SRT aus dem Jahre 1904 gelingt es ihm 
\emph{fast} (siehe unten), die allgemeine Lorentzinvarianz seiner 
elektrodynamischen Gleichungen zu zeigen und damit auch die 
exakte Abwesenheit jedweder Effekte einer Bewegung relativ 
zum "Ather~\cite{Lorentz:1904}. Nach einer 
Vorl"auferarbeit~\cite{Lorentz:1899} aus dem Jahre 1899 tauchen 
hier die exakten Lorentz-Transformationen (\ref{eq:LorentzTrans})
zum ersten Male im gr"o"seren Zusammenhang auf.\footnote{Lorentz 
l"a"st zun"achst noch konstante Skalentransformationen 
$\vec x'=l\cdot\vec x,\ t'=l\cdot t$ zu, von denen er aber 
sp"ater beweist, da"s sie keine Symmetrien des Kraftgesetzes f"ur 
die Ladungstr"ager sein k"onnen. Wir werden sie hier von 
vornherein weglassen.}

An dieser Stelle sollte folgendes herausgestellt werden:  
Der formal-mathematische Beweis der Lorentzinvarianz der 
Bewegungsgleichungen impliziert nur dann die physikalische Aussage 
der Unbeobachtbarkeit einer Bewegung relativ zum "Ather, wenn 
klargestellt ist, da"s die neuen Raumkoordinaten $\vec x'$ und 
Zeitkoordinaten $t'$ bez"uglich des bewegten Beobachters dieselbe 
physikalische Rolle spielen wie die ungestrichenen Koordinaten 
bez"uglich dem gegen den "Ather ruhenden Beobachter. Interpretiert 
man die Raum- und Zeitkoordinaten physikalisch durch r"aumliche 
Abst"ande, die mit wirklichen Ma"sst"aben (oder anderen physikalischen 
Methoden) gemessen werden bzw. Zeitintervallen, die von wirklichen 
Uhren angezeigt werden, so hat man damit eine implizite Aussage 
"uber das physikalische Verhalten von Ma"sst"aben und Uhren bei 
Bewegung relativ zum "Ather gemacht. 

Aus \cite{Lorentz:1904} geht nun eindeutig hervor, da"s 
Lorentz dieser Sachverhalt v"ollig klar war. Eingehend erkl"arte er, 
da"s die durch den Faktor $\gamma$ in (\ref{eq:LorentzTrans1}) 
bedingte "Anderung r"aumlicher Abst"ande nichts anderes sei
als seine alte Kontraktionshypothese, die f"ur alle Materiekomplexe 
gilt, sofern diese im Prinzip mit den vorliegenden Gleichungen 
beschrieben werden. Was Lorentz interessanterweise aber \emph{nicht}
sagt, ist, da"s auch alle Uhren, sofern sie auf elektromagnetischen
Prinzipien beruhen, die durch den Faktor $\gamma$ in 
(\ref{eq:LorentzTrans2}) bedingte Zeitdilatation anzeigen. Er mu"s 
es implizit annehmen, denn sonst -- das sei nochmals hervorgehoben -- 
gibt es keine logische Verbindung zwischen der mathematischen Aussage 
der Lorentzinvarianz und der physikalischen Aussage der 
Unbeobachtbarkeit von Translationsbewegungen gegen den "Ather. 
 
An dieser Stelle will ich auf die obige Bemerkung zur"uckkommen, 
da"s Lorentz die Invarianz seiner Gleichungen nur \emph{fast} 
gezeigt hat. Dieses "`fast"' bezieht sich auf folgenden 
kuriosen "`Fehler"' (a posteriori betrachtet) in 
$\cite{Lorentz:1904}$\footnote{Der Wiederabdruck der Lorentzschen 
Arbeit in \cite{LEM:RelPrinz} enth"alt eine sp"ater von Lorentz 
eingef"ugte Fu"snote, in der er diesen "'Fehler"' einr"aumt, ohne 
aber eine Bemerkung "uber die physikalische Bedeutung von $t'$ 
zu machen.}: Obwohl er die Lorentz-Transformationen exakt 
hinschreibt, setzt er \emph{unabh"angig} davon ein (falsches) 
Kompositionsgesetz f"ur die Geschwindigkeiten an, das nicht dem 
(Einsteinschen, vgl. (\ref{eq:GeschAddLor})) entspricht, welches 
zwingend aus den Transformationsformeln folgt. Letzteres gilt 
allerdings nur dann, wenn man die Zeit $t'$ im bewegten System 
wirklich als physikalische Zeit interpretiert, die auch zur Definition 
von Geschwindigkeiten (relativ zum bewegten System) herangezogen 
wird. Dies mag als ein Hinweis gewertet werden, da"s Lorentz 
seine "`Ortszeit"' $t'$ nicht konsequent physikalisch ernst nahm. 

Das zweite Hauptresultat der Lorentzschen Arbeit~\cite{Lorentz:1904}
betrifft die Geschwindigkeitsabh"angigheit der tr"agen 
elektromagnetischen Masse, f"ur die Lorentz genau die Beziehungen 
der SRT findet\footnote{Die Masse w"achst mit dem Faktor 
$\gamma=(1-v^2/c^2)^{-1/2}$ f"ur Beschleunigungen, die momentan 
senkrecht auf der Geschwindigkeit stehen ("`transversale Masse"')
und mit dem Faktor $\gamma^3$ f"ur Beschleunigungen parallel 
zur Geschwindigkeit ("`longitudinale Masse"').} Er argumentiert 
weiter, da"s diese Beziehungen f"ur alle tr"agen Massen gelten
sollten, also auch f"ur solche, die nicht elektromagnetischen Ursprungs 
sind. 

Zum Schlu"s sei erw"ahnt, da"s Einstein dem Abdruck seiner 
Originalarbeit zur SRT in \cite{LEM:RelPrinz} eine Fu"snote 
hinzugef"ugt hat, nach der ihm Lorentz' Arbeit \cite{Lorentz:1904} 
beim Abfassen seiner Arbeit nicht bekannt war.    

\subsection{Poincar\'e}
\label{sec:Poincare}
Poincar\'es Arbeiten, die ihn als Vordenker der SRT ausweisen,
zerfallen in auff"alliger Weise in zwei Gruppen: Die eine Gruppe 
enth"alt solche Arbeiten, die im technisch-mathematischen Stil 
elegant formuliert und damit nur dem Fachmann zug"anglich sind. 
Die Arbeiten der anderen Gruppe sind untechnisch geschrieben -- 
ohne Darstellung mathematischer Formeln -- und wenden sich 
an einen weiteren Kreis. Diese entstammen oft "offentlichen 
Vortr"agen und haben durchweg programmatischen Charakter von 
gro"ser Weitsicht. 

Ich m"ochte mich zuerst der ersten Gruppe zuwenden und dort nur 
die entscheidende Arbeit kurz zusammenfassen, die den Titel 
"`Sur la dynamique d'\'electron"' tr"agt. Sie existiert 
in zwei Versionen: einer Kurzversion~\cite{Poincare:1905},
die der Pariser Akademie der Wissenschaften am 5.\,Juni 1905 vorgelegt 
wurde und einer weit l"angeren Version~\cite{Poincare:1906}, die 
Poincar\'e im Juli~1905 abschlo"s und daraufhin bei der Zeitschrift
\emph{Rendiconti del Circolo Matematico di Palermo} einreichte, wo sie 1906 
publiziert wurde. Es ist am"usant zu bemerken, da"s das 
Einreichungsdatum der Einsteinschen Arbeit zur SRT der 30.\,Juni 1905 
war, also zwischen den beiden vorgenannten Terminen liegt.\footnote{%
Eine Korrelation ist allerdings so gut wie ausgeschlossen. Es gilt als 
sicher, da"s Einstein keine Kenntnis der ersten Poincar\'eschen Arbeit 
hatte.}

Es ist in der Tat beeindruckend, wie viele der uns heute aus der SRT
bekannten Ergebnisse sich bereits in der zweiten, ausf"uhrlicheren 
Arbeit Poincar\'es finden. Um einen ersten Eindruck zu vermitteln, 
seien nur einige der Poincar\'eschen Ergebnisse zusammengefa"st: 
\begin{itemize}
\item
Aussprechen des Relativit"atsprinzips in voller Allgemeinheit.
\\[-7mm]
\item
Korrekter Beweis der Lorentzinvarianz der Lorentz-Maxwell-Gleichungen.
Herleiten derselben aus einem invarianten Wirkungsprinzip. 
\\[-7mm]
\item
Beweis, da"s die Lorentz-Transformationen zusammen mit den 
r"aumlichen Drehungen eine Gruppe bilden, die er 
\emph{Lorentzgruppe} nennt.
\\[-7mm]
\item
Beweis der Lorentzinvarianz des Linienelements 
$c^2(\Delta t)^2-\Delta\vec x\cdot\Delta\vec x$.
\\[-7mm]
\item
Ableitung des richtigen Kompositionsgesetzes f"ur 
Geschwindigkeiten. 
\end{itemize}  
Ein erheblicher Teil dieser Arbeit ist der Frage gewidmet, 
ob man die FitzGerald-Lorentz-Kontraktion (Poincar\'e nennt 
sie nur \emph{Lorentz-Kontraktion}, was sich seither eingeb"urgert hat) 
\emph{dynamisch} aus einem elektromagnetischen Modell des Elektrons 
ableiten kann. 
In diesem Zusammenhang entwickelt Poincar\'e sein klassisches 
Modell des Elektrons, in dem die Annahme eines Drucks (von noch 
unbekannter Natur) die Explosion der 
gleichnamig geladenen (und daher sich gegenseitig absto"senden) 
Teile verhindern. Zum Schlu"s geht er noch der Frage nach,
wie man die Gravitation durch eine lorentzinvariante Theorie 
beschreiben kann. In diesem Zusammenhang spricht er auch von 
"`Gravitationswellen"'. Erw"ahnt werden mu"s, da"s Poincar\'e 
in \emph{dieser} Arbeit an keiner Stelle von Definition der 
Gleichzeitigkeit oder der physikalischen Bedeutung der 
transformierten Zeit $t'$ spricht. Somit bleibt auch bei ihm 
(wie zuvor bei Lorentz) unklar, welche logische Verbindung 
zwischen der formal-mathematischen Lorentzinvarianz der 
zugrundegelegten Gleichung einerseits und der G"ultigkeit des 
physikalischen Relativit"atsprinzips andererseits besteht. 

Diese L"ucke wird andeutungsweise geschlossen, wenn wir 
uns den anderen, rein programmatischen Arbeiten Poincar\'es 
zuwenden. Von diesen waren wahrscheinlich nur diejenigen Einstein 
bekannt, die in Poincar\'es Aufsatzsammlung "`Wissenschaft und 
Hypothese"' enthalten sind. Dieser Sammelband erschien 1902 in 
franz"osischer und 1904 in deutscher Sprache und wurde in der 
"`Akademie Olympia"' von Einstein und seinen Freunden 
Maurice Solovine und Conrad Habicht intensiv gelesen (im 
franz"osischen Original), wie Solovine berichtet (zitiert in
\cite{Pais:Einstein}, p.\,133): 
"`Dieses Buch beeindruckte uns tief und hielt uns wochenlang 
gefangen"'. 

Darin besonders interessiert haben d"urfte Einstein die allgemeine 
Diskussion der Lorentzschen Arbeiten zum Relativit"atsprinzip. 
Poincar\'e erw"ahnt den Lorentzschen Beweis 1.\,Ordnung (in $v/c$) 
von 1895 und dessen Verallgemeinerung auf die quadratische 
Ordnung vermittels der Kontraktionshypothese. Er kritisiert den
ad-hoc-Charakter dieser Hypothese u.a. mit den Worten "`an Hypothesen 
ist niemals Mangel"'~(\cite{Poincare:WuH}, p.\,173) und dr"uckt seinen 
Unmut dar"uber aus, Ordnung f"ur Ordnung nachrechnen zu m"ussen, 
da"s das von ihm als fundamental erachtete Relativit"atsprinzip 
wie durch Zufall auch tats"achlich gilt. Auf die rhetorisch von ihm 
gestellte Frage, ob es sich denn dabei lediglich um einen 
gl"ucklichen Umstand handeln k"onne, gibt er selbst die Antwort 
(\cite{Poincare:WuH}, p.\,173): 
\begin{quote}
"`Nein, man mu"s f"ur die einen wie die anderen [Ordnungen] 
dieselbe Erkl"arung finden, und dann dr"angt uns alles darauf 
hin, zu erw"agen, da"s diese Erkl"arung gleicherweise f"ur die 
h"ohere Ordnung gelten w"urde und da"s die gegenseitige 
Zerst"orung dieser Glieder eine strenge und absolute ist."'
\end{quote}

In einem anderen Aufsatz dieser Sammlung schreibt Poincar\'e 
"uber Zeit und Zeitmessung folgendes (\cite{Poincare:WuH}, p.\,92):  
\begin{quote}
"`Es gibt keine absolute Zeit; wenn man sagt, da"s zwei 
Zeiten gleich sind, so ist das eine Behauptung, welche 
an sich keinen Sinn hat und welche einen solchen nur durch 
"Ubereinkommen erhalten kann. 
Wir haben nicht nur keinerlei direkte  Anschauung von der 
Gleichheit zweier Zeiten, sondern wir haben nicht einmal 
diejenige von der Gleichzeitigkeit zweier Ereignisse, 
welche auf verschiedenen Schaupl"atzen vor sich gehen; das 
habe ich in einem Aufsatze unter dem Titel: la Mesure du 
temps [in einer Fu"snote wird die Referenz gegeben] dargelegt."'
\end{quote}
An dieser Stelle wird es nun wirklich interessant! 
Denn schaut man sich diese Referenz~\cite{Poincare:1898} 
an\footnote{Es gibt meines Wissens keinerlei eindeutigen Hinweis 
darauf, ob Einstein diese Schrift gelesen hat oder nicht.}, so 
findet man darin im Zusammenhang mit der Frage, wie Gleichzeitigkeit 
zu definieren ist, auch eine Er"orterung, wie ein Astronom die 
Lichtgeschwindigkeit mi"st (denn diese mu"s er ja kennen, um 
Ereignissen, die fern von ihm stattgefunden haben, eine Zeit 
zuzuordnen).  Poincar\'e  schreibt: 
\begin{quote}
"`Er [der Astronom] hat zun"achst angenommen, da"s das 
Licht eine konstante Geschwindigkeit hat und besonders, 
da"s seine Geschwindigkeit nach allen Richtungen die 
gleiche ist. Das ist ein Postulat, ohne das keine 
Messung dieser Geschwindigkeit versucht werden k"onnte.
Dies Postulat wird nie durch die Erfahrung unmittelbar
best"atigt werden k"onnen; es k"onnte aber durch sie 
widerlegt werden, wenn die Resultate verschiedener 
Messungen nicht "ubereinstimmend w"aren."'  
\end{quote}
Die Vorschrift, die Uhren so zu synchronisieren, da"s die 
Lichtausbreitung isotrop, d.h. in allen Richtungen mit der 
gleichen Geschwindigkeit erfolgt, ist nichts anderes als 
die heute so genannte "`Einsteinsche Synchronisationsvorschrift"'. 
Was Poincar\'e richtig herausstellt, ist 1) die Notwendigkeit
einer Konvention betreffend die Uhrensynchronisation (es mu"s 
allerdings nicht notwendigerweise die der Isotropie sein), 
um von der Lichtgeschwindigkeit entlang einer Strecke zwischen 
\emph{unterschiedlichen} Raumpunkten zu sprechen und 2) die 
Nicht-Trivialit"at der Tatsache, da"s die erw"ahnte Konvention 
"uberhaupt widerspruchsfrei m"oglich ist. Um letzteres zu 
erm"oglichen, m"ussen n"amlich bestimmte naturgesetzliche 
Voraussetzungen vorliegen.  Poincar\'e endet seinen Artikel 
mit den Worten:  
\begin{quote}
"`Die Gleichzeitigkeit zweier Ereignisse oder ihre 
Aufeinanderfolge und die Gleichheit zweier Zeitr"aume 
m"ussen derart definiert werden, da"s der Wortlaut der 
Naturgesetze so einfach als m"oglich wird.      
Mit anderen Worten, alle diese Regeln, alle diese 
Definitionen sind nur die Fr"uchte eines unbewu"sten 
Opportunismus."'
\end{quote}

Es ist kein Zufall, da"s sich der Mathematiker Poincar\'e mit 
solch "`praktischen"' Problemen wie der Frage der operationalen
Synchronisation von Uhren besch"aftigte. Seit 1893 war er Mitglied 
des \emph{Bureau des Longitudes} (und 1899 ihr Pr"asident), wo er  
aktiv mit Problemen telegraphischer Synchronisationsverfahren 
konfrontiert war. Die dabei gemachten Erfahrungen spiegeln sich 
direkt in seinen programmatischen Schriften wider.\footnote{%
Die diesbez"uglichen technologischen Fragestellungen der 
damaligen Zeit sowie ihre Relevanz f"ur die Entstehungsgeschichte 
der SRT, werden eingehend in \cite{Galison} beleuchtet.}

Das trifft vor allem f"ur eine Serie von Vortr"agen zu, die 
Poincar\'e 1904 auf dem \emph{Congress of Arts and Science} in St.\,Louis 
hielt (abgedruckt als Kap.\,7-9 in \cite{Poincare:WdW}). 
Dort gibt er eine noch explizitere Schilderung der 
(Einsteinschen) Synchronisationsvorschrift f"ur Uhren
(\cite{Poincare:WdW}, p.\,141-142): 
\begin{quote}
"`Die allerscharfsinnigste Idee ist die der lokalen Zeit.
Denken wir uns zwei Beobachter, die ihre Uhren nach 
optischen Signalen regulieren wollen. Sie tauschen Signale;
da sie aber wissen, da"s die "Ubertragung des Lichtes nicht 
augenblicklich geschieht, m"ussen sie darauf bedacht sein, 
sie zu kreuzen. Wenn die Station $B$ das Signal der Station 
$A$ bemerkt, darf ihre Uhr nicht die gleiche Zeit 
zeigen wie die der Station $A$ im Augenblick der Aussendung
des Signals, sondern eine Zeit, die um einen konstanten, 
die Dauer der "Ubertragung bedeutenden Zeitraum sp"ater 
ist. Nehmen wir zum Beispiel an, da"s die Station $A$ ihr 
Signal abgibt, wenn ihre Uhr `Null' zeigt und die Station 
$B$ es bemerkt, wenn ihre Uhr $t$ zeigt. Die Uhren sind 
gerichtet, wenn die [der Zeitspanne] $t$ gleiche Verz"ogerung 
die Dauer der "Ubertragung bedeutet, und um es zu erproben, 
sendet die Station $B$ ihrerseits ein Signal, wenn ihre Uhr 
auf Null steht, und die Station $A$ mu"s es nun bemerken, 
wenn ihre Uhr $t$ zeigt. Dann sind die Uhren reguliert."'
\end{quote}
Mit "`lokaler Zeit"' bezieht er sich auf die "`Ortszeit"'
Lorentz'. Wieder gibt er 
exakt die Definition der Einsteinschen Synchronisationsvorschrift,
die zu der oben gegebenen vollst"andig "aquivalent ist. 
Man k"onnte meinen, da"s Poincar\'e jetzt das bei Lorentz noch 
fehlende Element einer konsequenten operationalistischen 
Interpretation der "`Ortszeit"' in H"anden h"alt. Doch, und das 
ist vielleicht die eigentliche "Uberraschung, geht er diesen 
Schritt nicht.\footnote{In \cite{Damour:Poincare} wird sogar 
argumentiert, da"s dies nicht einmal f"ur die Zeit nach 1905 
der Fall war.} Die sich anschlie"senden Worte offenbaren deutlich, 
da"s er zu dieser Zeit konzeptuell immer noch n"aher bei Lorentz 
als bei Einstein stand:   
\begin{quote}
"`Die auf diese Weise gerichteten Uhren zeigen also nicht die 
wahre Zeit; was sie zeigen, k"onnte man lokale Zeit nennen;
die eine wird gegen die andere nachgehen. Es liegt aber 
nichts daran, da wir kein Mittel haben, es zu bemerken. 
Alle Erscheinungen, die zum Beispiel in $A$ entstehen, 
versp"aten sich, aber sie tun es alle gleichm"a"sig, und der 
Beobachter wird es nicht bemerken, weil seine Uhr nachgeht; 
also hat er, wie es das Prinzip der Relativit"at verlangt, 
gar kein Mittel zu wissen, ob er in absoluter Ruhe oder in 
Bewegung ist."'      
\end{quote}  
Wie bei Lorentz herrscht bei Poincar\'e die Vorstellung, das 
Relativit"atsprinzip g"alte in der Elektrodynamik als Resultat 
einer (wenn auch sehr fundamental angesiedelten) 
\emph{dynamischen Konspiration}. 
Es gibt den "Ather und damit ein ausgezeichnetes Bezugssystem, 
es gibt die wahre Zeit und damit die absolute Gleichzeitigkeit, 
doch konspirieren die dynamischen Naturgesetze (jedenfalls die der 
Elektrodynamik) gerade so, da"s uns diese Strukturen unsichtbar 
bleiben m"ussen. 

Etwas pointiert ausgedr"uckt bekommt man den Eindruck, da"s auf 
erkenntnistheoretischem Niveau das Relativit"atsprinzip f"ur 
Poincar\'e letztlich den Status einer von der Natur "uberaus 
raffiniert eingerichteten optischen T"auschung besitzt, obwohl er 
es selber in die Reihe der fundamentalen physikalischen Prinzipien 
einreiht~(\cite{Poincare:WdW}, p.\,140). Dazu pa"st, da"s Poincar\'e 
die Erf"ullung des (physikalischen) Relativit"atsprinzips logisch 
keinesfalls gleichsetzt mit der formal-mathematischen Eigenschaft 
der Lorentzinvarianz. In dieser Beziehung ist er deutlicher als 
Lorentz (siehe oben). Im direkten Anschlu"s an die eben zitierten 
"Au"serungen f"ahrt er fort: 
\begin{quote}
"`Das gen"ugt leider noch nicht [um die Erf"ullung des 
Relativit"atsprinzips zu gew"ahrleisten], und man braucht 
erg"anzende Hypothesen; man mu"s annehmen, da"s die in 
Bewegung befindlichen K"orper eine gleichm"a"sige Kontraktion 
in der Richtung der Bewegung erleiden."'   
\end{quote}
F"ur Poincar\'e ist die allgemeine Kontraktionshypothese also 
noch keine logische Folge der Lorentzinvarianz der elektromagnetischen 
Grundgleichungen, womit er nat"urlich recht hat! Denn diese 
Implikation g"alte ja nur dann, wenn bereits gezeigt w"are, da"s 
Aufbau und Geometrie der betreffenden K"orper allein durch die 
elektromagnetischen Grundgleichungen bestimmt ist, was zu diesem 
Zeitpunkt eine unbewiesene Hypothese war.\footnote{Die bereits 
erw"ahnte Poincar\'esche Theorie des Elektrons, die er in 
\cite{Poincare:1906} entwickelte,  war f"ur ihn ein erster Schritt 
in diese Richtung.}  Alternativ m"u"ste man beweisen (oder zumindest 
postulieren), da"s eventuell nicht elektromagnetische Kr"afte, die 
beim Aufbau der K"orper eine Rolle spielen k"onnten, ebenfalls durch 
lorentzinvariante Grundgleichungen bestimmt sind. 

Zusammenfassend kann man wohl sagen, das Poincar\'e der SRT \emph{sehr} 
nahe kam, den eigentlich entscheidenden Schritt aber nicht vollzog,
auch nicht -- nach allem was die zitierten Quellen hergeben -- gedanklich.
Er verstarb 1912 und hat sich seltsamerweise nie "offentlich zu 
Einstein und/oder der SRT ge"au"sert (mehr dazu in \cite{Pais:Einstein}). 
Darin unterscheidet sich Lorentz, dem die Radikalit"at der Einsteinschen 
Vorgehensweise nicht sympathisch war, was er auch "offentlich aussprach.  
So z.B. in einer Vorlesung aus dem Jahre 1913 (\cite{Lorentz:1914}, p.\,23): 
\begin{quote}
"`Einstein sagt kurz und gut, da"s alle soeben genannten Fragen
[nach absoluter L"ange und absoluter Gleichzeitigkeit] keinen Sinn 
haben. Er kommt denn auch zu einem "<Abschaffen"> des "Athers. [...]
Es ist gewi"s merkw"urdig, da"s diese Relativit"atsbegriffe, auch 
was die Zeit betrifft, so schnell Eingang gefunden haben. [...] 
Soweit es den Vortragenden betrifft, findet er die 
"altere Deutung zufriedenstellender, wonach der "Ather 
eine gewisse Substantialit"at besitzt, Raum und Zeit 
streng voneinander trennbar sind und Gleichzeitigkeit 
ohne Einschr"ankungen definiert werden kann. 
In Bezug auf den letzten Punkt kann man vielleicht 
an unsere F"ahigkeit appellieren, sich beliebig gro"se 
Geschwindigkeiten vorzustellen. Damit kommt man dem 
Begriff absoluter Gleichzeitigkeit sehr nahe. [..]
Schlie"slich ist anzumerken, da"s die k"uhne Behauptung
"uber die Unbeobachtbarkeit von "Uberlichtgeschwindigkeiten
eine hypothetische Einschr"ankung des uns Zug"anglichen 
beinhaltet, die nicht ohne Zur"uckhaltung anerkannt werden kann."' 
\end{quote}

\section{Einstein: Zeit ist, was man auf der Uhr abliest}
\label{sec:Ausblick}
In seiner Originalarbeit \cite{Einstein:SRT} zur SRT kehrt 
Einstein die logische Hierarchie der Lorentz-Poincar\'eschen 
Argumentation gerade um: Bei ihm  ist die G"ultigkeit des 
Relativit"atsprinzips nicht Folge einer "`Konspiration"' zwischen 
dynamischen Gesetzen, die es m"uhsam aufzudecken gilt, sondern 
selbst prim"ares physikalisches Prinzip, das allen dynamischen
Gesetzen zugrundeliegt.\footnote{Mit Ausnahme der Gravitation gilt 
dies heute noch f"ur die drei verbleibenden, fundamentalen 
Wechselwirkungen: die starke, die schwache und die elektromagnetische.} 
Das zweite Postulat, auf dem die Einsteinsche Arbeit aufbaut, 
ist das der \emph{Konstanz der Lichtgeschwindigkeit}. Um dies pr"azise zu 
formulieren, mu"s man sich erst wieder vergegenw"artigen, da"s von 
einer Geschwindigkeit entlang eines nicht geschlossenen Weges erst 
dann gesprochen werden kann, \emph{nachdem} man sich "uber ein 
Verfahren zur Synchronisation der Uhren an den Wegenden geeinigt 
hat. Das zweite Postulat m"ochte ich dann so aussprechen:\footnote{%
Man beachte, da"s dieses Postulat nicht nur dem Relativit"atsprinzip 
widersprechen w"urde, sondern auch zu einer Trivialit"at verk"ame, 
wenn man nicht forderte, da"s die Synchronisationsvorschrift in 
allen Inertialsystemen die gleiche w"are.}

\vspace{0.5cm}
\noindent
\textbf{Postulat der Konstanz der Lichtgeschwindigkeit:}
\emph{Es gibt eine in allen Inertialsystemen gleichlautende Vorschrift 
zur gegenseitigen Synchronisation der im jeweiligen Inertialsystem 
ruhenden Uhren, so da"s die Lichtgeschwindigkeit f"ur alle 
Inertialsysteme und in allen Richtungen den gleichen Wert $c$ 
besitzt.} 

\vspace{0.5cm}
\noindent
Der aus dem klassischen Kompositionsgesetz f"ur Geschwindigkeiten 
(\ref{eq:GeschAddGal}) sich ergebende Widerspruch zwischen Konstanz 
der Lichtgeschwindigkeit und Relativit"atsprinzip verschwindet,
da bez"uglich der so synchronisierten Uhren eben nicht die 
Galilei-Transformationen  (\ref{eq:GalileiTrans}), sondern die 
Lorentz-Transformationen (\ref{eq:LorentzTrans}) gelten. Aus 
diesen folgt f"ur die Komposition gleichgerichteter 
Geschwindigkeiten statt $w=w'+v$ das Gesetz: 
\begin{equation}
\label{eq:GeschAddLor}
w=\frac{w'+v}{1+\frac{w'v}{c^2}}\,.
\end{equation}
Aus diesem folgt, da"s die Komposition zweier 
Unterlichtgeschwindigkeiten immer wieder eine 
Unterlichtgeschwindigkeit ist und da"s die Komposition einer 
Unterlichtgeschwindigkeit und der Lichtgeschwindigkeit immer gleich 
der Lichtgeschwindigkeit ist. 

Als einfache Anwendung zeigen wir die Ableitung des Fresnelschen 
Mitf"uhrungsgesetzes (\ref{eq:Fizeau}): Innerhalb eines K"orpers 
vom Brechungsindex $n$ bewege sich ein Lichtstrahl mit der 
Geschwindigkeit $w'=c/n$. Der K"orper bewege sich mit der 
Geschwindigkeit $v$ relativ zum Laborsystem in die gleiche Richtung.
Dann hat nach (\ref{eq:GeschAddLor}) der Lichtstrahl relativ zum 
Laborsystem die Geschwindigkeit 
\begin{equation}
\label{eq:FresnelSRT}
w=\frac{(c/n)+v}{1+\frac{v}{nc}}\approx
\left(\frac{c}{n}+v\right)\left(1-\frac{v}{nc}\right)\approx
\frac{c}{n}+\underbrace{v\cdot\bigl(1-n^{-2}\bigr)}_{=:\,\Delta v}\,.
\end{equation}
Dabei bedeutet $\approx$ Gleichheit bis auf Terme der Ordung $v/c$
(im Fizeauschen Experiment war $v/c$ von der Gr"o"senordung $10^{-8}$).     

F"ur Einstein ist es selbstverst"andlich, da"s eine bewegte Uhr
die transformierte Zeit $t'$ anzeigt, denn das Gesetz, das den 
Gang der Uhr regelt, soll ja ebenfalls dem Relativit"atspostulat 
gen"ugen (d.h. lorentzinvariant sein). So bemerkt er gleich zu 
Beginn seiner Arbeit:
\begin{quote}     
"`Es k"onnte scheinen, da"s alle die Definitionen der "<Zeit"> 
betreffenden Schwierigkeiten dadurch "uberwunden werden k"onnen,
da"s ich an Stelle der "<Zeit"> die "<Stellung des kleinen Zeigers
meiner Uhr"> setze."' 
\end{quote}
Popul"ar wird diese "Au"serung oft durch das in der "Uberschrift 
dieses Abschnitts angegebene Diktum wiedergegeben. 

Es wird Inhalt des Beitrages von Herrn Dr.\,Hunziker sein, genauer 
auf die Kinematik der SRT einzugehen. Ich m"ochte daher zum Abschlu"s 
nur noch einige Bemerkungen zum Verh"altnis der Einsteinschen zur 
Lorentz-Poincar\'eschen Sichtweise machen: Die entscheidende Leistung
Einsteins, die seine Behandlung des Relativit"atsproblems "uber die 
Lorentz' und Poincar\'es heraushebt, liegt in der Neuformulierung 
und Nutzbarmachung eines Zeitkonzepts, das dem operationalen Vorgehen 
des Physikers unmittelbar angepa"st ist. Da"s dieses Zeitkonzept auch 
einen revolution"aren Kern beinhaltet, der direkt mit den mehr 
spektakul"aren Folgerungen der SRT verbunden ist (Uhrenparadoxon, 
Masse-Energie-"Aquivalenz etc.), ist an dieser Stelle eher 
nebens"achlich. Zun"achst hat Einstein seinen Vorarbeitern Lorentz 
und Poincar\'e ja nur das voraus, da"s er auf der Basis des 
allgemeing"ultig \emph{postulierten} Relativit"atsprinzips die 
gleichberechtigte physikalische Relevanz der Lorentzschen "`Ortszeit"' 
$t'$ klar macht und die nur durch Vorurteile gest"utzte Sonderrolle 
der absoluten ("`wahren"') Zeit $t$ beseitigt. 

Oft wir der Unterschied der Auffassungen zwischen Lorentz und 
Poincar\'e auf der einen Seiten und Einstein auf der anderen anhand 
der Lorentz-Kontraktion so erl"autert: f"ur Lorentz und Poincar\'e 
ist die Kontraktion ein \emph{dynamischer} Proze"s, f"ur Einstein dagegen 
ein \emph{kinematischer }. Was ist damit gemeint? Dazu zun"achst die 
Bemerkung, da"s diese Begriffe physikalisch nicht als prinzipiell 
disjunkt zu verstehen sind. Kinematischen Aussagen beruhen letztlich 
immer auch auf Annahmen "uber die Dynamik der Systeme, die man 
physikalisch zur Realisierung metrischer und deshalb auch 
kinematischer Begriffe verwendet. So h"angt z.B. das Verhalten einer 
Uhr oder eines Ma"sstabes in Raum und Zeit von deren dynamischen 
Gesetzen ab. Das ist auch in der SRT so, wo stets implizit 
vorausgesetzt wird, da"s Ma"sst"abe und Uhren selbst 
Lorentzinvarianten dynamischen Gesetzen unterliegen. Mit kinematisch 
kennzeichnet man im allgemeinen eine Aussage, die zwar eigentlich 
auch die dynamischen Gesetze physikalischer Systeme angeht, aber 
solch eine Allgemeing"ultigkeit besitzt, da"s sie von den 
spezifischen dynamischen Unterschieden aller in Betracht 
gezogenen Systeme g"anzlich unabh"angig ist. 
 
Als letzte Bemerkung folgendes: Oft ensteht der falsche Eindruck, 
als seien die neuen und ungew"ohnlichen Aussagen der SRT s"amtlich 
eine Folge der besonderen Wahl f"ur die (Einsteinsche) 
Synchronisationsvorschrift und in diesem Sinne nicht real, weil 
konventionsabh"angig. Dies ist aber so nicht richtig. Nat"urlich 
gibt es, wie in jedem physikalischen Begriffssystem, auch in der SRT 
Aussagen, die konventionsabh"angig sind, doch betrifft dies niemals 
das ganze System. Beobachtbare Gr"o"sen k"onnen letztlich immer 
in konventions\emph{unabh"angiger} Weise ausgedr"uckt werden. So mi"st 
man z.B. im Michelson-Morley-Experiment die Differenz der 
Lichtlaufzeiten auf \emph{geschlossenen} Wegen (d.h. mit \emph{einer} 
Uhr am Beobachterort), so da"s die Wahl der Synchronisation f"ur 
die Beschreibung dieses Experiments v"ollig irrelevant ist, was 
leider nicht immer in gen"ugender Klarheit gesagt wird. 
Nicht konventionalistisch sondern tats"achlich von naturgesetzlicher 
Art ist z.B. auch die Aussage, da"s die Einsteinsche 
Synchronisationsvorschrift "uberhaupt m"oglich ist. 

Erst durch die Einsteinsche Formulierung bekommt das 
Relativit"atsprinzip die enorme heuristische\footnote{Das 
griechische Verb `heuriskein' bedeutet `finden' oder `auffinden'. 
Die Heuristik ist die Kunst des Findens von Neuem. Die Forderung 
des Relativit"atsprinzips in Form der Lorentzinvarianz ist 
eine sehr effektive Konstruktionshilfe beim Aufstellen neuer 
dynamischer Gesetze.} Kraft, die f"ur die Entwicklung der 
Physik des 20.\,Jahrhunderts eine gro"se Rolle gespielt hat.
Technisch erm"oglicht wurde sie durch die vision"aren 
mathematischen Gedanken \cite{Minkowski:1909} Hermann Minkowskis 
(1864-1909), die dem Einsteinschen Programm erst die n"otige 
formale Durchschlagskraft verliehen. Grundvoraussetzung 
der ganzen Entwicklung bleibt aber Einsteins Vorurteilslosigkeit, 
gepaart mit seinem vielfach bewiesenen Mut, eine Sache einmal ganz 
anders anzugehen. Teil daran hat sicher auch sein Aufenthalt an der 
Aarauer Kantonsschule, deren spezielle Atmosph"are ein Mitsch"uler 
Einsteins, Hans Byland (1878-1949), einmal wie folgt charakterisierte
\cite{Byland:1928}: 
\newpage
\begin{quote}
"`An der Aargauischen Kantonsschule wehte in den 90er Jahren ein
scharfer Wind der Skepsis, worauf schon die Tatsache hindeutet,
da"s aus meiner Klasse, so wenig als aus den zwei n"achsten, kein
Theologe hervorging. In diese Atmosph"are pa"ste der kecke 
Schwabe [Einstein] nicht "ubel"'.
\end{quote}
Doch niemand formulierte das charmanter als Einstein 
in seiner "`Autobiographischen Skizze"'
(\cite{Seelig:HelleZeit}, p.\,9-10): 
\begin{quote}
"`Diese Schule hat durch ihren liberalen Geist und durch den schlichten 
Ernst der auf keinerlei "au"serliche Autorit"at sich st"utzenden
Lehrer einen unverge"slichen Eindruck in mir hinterlassen;
durch Vergleich mit sechs Jahren Schulung an einem deutschen,
autorit"ar gef"uhrten Gymnasium wurde mir eindringlich bewu"st, 
wie sehr die Erziehung zu freiem Handeln und Selbstverantwortlichkeit 
jener Erziehung "uberlegen ist, die sich auf Drill, "au"sere 
Autorit"at und Ehrgeiz st"utzt. 
Echte Demokratie ist kein leerer Wahn"'.
\end{quote}

%

\section{Zeittafel}
\label{sec:Zeittafel}
Die folgende, stichwortartige Aufstellung gibt nochmals eine 
chronologische "Ubersicht "uber einige historische Eckdaten vor 
1905, die mit der Geschichte der SRT in Zusammenhang stehen. 
%
\begin{itemize}
\item
\emph{R{\o}mer 1675/76:}
Messung der Lichtgeschwindigkeit anhand der Verdunkelungen 
des Jupitermondes Io. Der gemessene Wert lag etwa 25\%
unterhalb des heute bekannten, genauen Wertes. 
\item
\emph{Bradley 1728:}
Entdeckung der stellaren Aberration bei dem Versuch, die
Fixsternparalaxen zu messen (letzteres gelang erst Bessel 
im Jahre 1838). 
\item
\emph{Young und Fresnel Anfang des 19.\,Jahrh.:} 
Wellentheorie des Lichtes. Analogie mit elastischen 
Schwingungen eines hypothetischen Mediums "`"Ather"', 
der alle Materie durchdringt. Dieser wurde anf"anglich 
als absolut ruhend (d.h von bewegter Materie nicht 
"`mitgeschleppt"') angesehen, wodurch Erkl"arung der 
stellaren Aberration m"oglich wird. 
\item
\emph{Malus und Brewster 1808-15:}
Entdeckung der Polarisierbarkeit von Lichtwellen, 
so da"s diese im "Atherbild nur Transversalwellen sein 
k"onnen. Damit ist der "Ather eher einem festen K"orper 
denn einer Fl"ussigkeit analog. 
\item
\emph{Arago 1810:}
Erster Versuch, eine Bewegung der Erde relativ zu einem 
ausgezeichneten Bezugssystem (z.B. Ruhesystem des "Athers) 
mit Hilfe optischer Experimente festzustellen, mit negativem 
Resultat~\cite{Arago:1810}.  
\item
\emph{Fresnel 1818:}
Stellt fest, da"s der negative Ausgang des Experiments von Arago 
(1810) durch die Hypothese erkl"art werden kann, da"s das Licht 
im Inneren eines durchsichtigen bewegten K"orpers, dessen Material
den Brechungsindex $n$ besitzt, relativ zum "Ather um den Bruchteil 
$f=(1-n^{-2})$ der K"orpergeschwindigkeit mitgeschleppt wird
($f$ ist bis heute als "`Fresnelscher Mitf"uhrungskoeffizient"' 
bekannt.). Er deutet dies im Rahmen einer "Athertheorie als 
Mitschleppung des "Athers \cite{Fresnel:1818}.   
\item
\emph{Stokes 1846-48:}
Motiviert durch die Analogie mit einem festen K"orper, propagiert 
Stokes im Gegensatz zu Fresnel einen "Ather, der durch 
Bewegung von Materie vollst"andig mitgeschleppt wird, was 
allerdings im Gegensatz zur stellaren Aberration zu stehen scheint. 
\item
\emph{Fizeau und Foucault 1849-50:}
Terrestrische Pr"azisionsmessungen der Lichtgeschwindigkeit
mit Werten n"aher als $5\%$ (1849 Fizeau mit rotierendem Zahnrad) 
und n"aher als $0{,}5\%$ (1850 Foucault mit rotierenden Spiegeln). 
\item
\emph{Fizeau 1851:} 
Mi"st Lichtausbreitung im str"omenden Wasser und best"atigt 
(f"ur dieses Medium) den Fresnelschen 
Mitf"uhrungskoeffizienten~\cite{Fizeau:1851}. 
\item
\emph{Maxwell 1864-65:} 
Stellt der Royal Society die Gleichungen (Maxwell-Gleichungen) seiner 
umfassenden Nahwirkungstheorie des Elektromagnetismus vor. 
Sagt die Existenz elektromagnetischer Wellen voraus, deren 
Geschwindigkeit er mit $310\,000\ \text{Km/s}$ angibt. Aus der 
N"ahe dieses Wertes zur Lichtgeschwindigkeit wagt er die Hypothese
("`we have strong reasons to conclude"'), da"s Lichtwellen (und 
W"armestrahlung) elektromagnetischer Natur sind. Seine ber"uhmte 
\emph{Treatise on Electricity and Magnetism} erscheint 1873 bei Oxford
University Press.
\item
\emph{Maxwell 1879:} 
In seinem Artikel "`Ether"' in der Encyclopaedia Britannica (9.th ed.) 
schl"agt er die mehrfache Wiederholung des R{\o}merschen Experiments "uber 
einen Abstand von mindestens 6 Jahren (halbe Umlaufzeit des Jupiter 
um die Sonne) vor, um damit die (m"oglicherweise sehr gro"se) 
Relativgeschwindigkeit des Sonnensystems zum "Ather zu messen
(Anisotropie der Lichtgeschwindigkeit). Wohl haupts"achlich wegen 
der Ungenauigkeit der R{\o}merschen Methode kommt es aber nicht zur 
Ausf"uhrung, beeinflu"st aber sp"atere Experimente.  
\item
\emph{Airy 1871:}
F"ullt Fernrohr mit Wasser und stellt fest, da"s die stellare 
Aberration davon nicht beeinflu"st (schw"acher) wird, das Wasser 
also keine merkliche "Athermitf"uhrung bewirkt.  
\item
\emph{Michelson 1881:} 
Erster Anlauf eines optischen Interferenzversuchs zum Nachweis 
einer Bewegung der Erde relativ zum "Ather; diesmal noch 
mit ungen"ugender Genauigkeit. 
\item
\emph{Michelson und Morley 1886:}
Pr"azisere Messungen als die Fizeaus (1851) best"atigen erneut 
den Fresnel-Koeffizienten~\cite{MichelsonMorley:1886}. 
\item
\emph{Michelson und Morley 1887:} 
Zweiter Anlauf eines optischen Interferenzversuchs zum Nachweis 
einer Bewegung der Erde relativ zum "Ather; diesmal mit 
mehr als ausreichender Genauigkeit -- Ausgang 
negativ~\cite{MichelsonMorley:1887}.
\item
\emph{Voigt 1887:} 
Studiert Wellengleichungen und stellt mathematische Invarianz 
unter Lorentz-Transformationen (erst 1905 von Poincar\'e so 
bezeichnet) fest~\cite{Voigt:1887}.  
\item
\emph{FitzGerald 1889:}
Erste Formulierung der Kontraktionshypothese zur Erkl"arung des 
negativen Ausgangs des Michelson-Morley-Experiments~\cite{FitzGerald:1889}.  
\item
\emph{Hertz 1890:} 
Erste vollst"andige Ausarbeitung der Maxwellschen Elektrodynamik 
f"ur bewegte K"orper unter der Arbeitshypothese eines vollst"andig 
mitgef"uhrten "Athers (Stokessche Annahme)~\cite{Hertz:1890b}. 
\item
\emph{Lorentz 1892:}
Erste Ver"offentlichung zur Elektronentheorie und erste 
(von FitzGerald unabh"angige) Erw"ahnung der 
Kontraktionshypothese als M"oglichkeit zur Erkl"arung des 
negativen Ausgangs des Michelson-Morley-Experiments.
\item
\emph{Lorentz 1894:}
Lorentz schreibt an FitzGerald, da"s er von seiner 
Kontraktionshypothese durch eine Arbeit von O.\,Lodge aus 
dem Jahr 1893 erfahren habe. 
\item
\emph{R"ontgen 1895:} 
Experiment mit Bewegung eines Isolators im elektrischen Feld 
widerlegt Hertzsche Theorie (1890) der Elektrodynamik in bewegten 
Medien, die von der Hypothese eines vollst"andig mitgef"uhrten 
"Athers (Stokes) ausgeht. In modifizierter Form wiederholt 
von \emph{Eichenwald 1903}; f"ur Magnetfelder von \emph{Wilson 1905}. 
\item
\emph{Lorentz 1895:} Umfassende Monographie \cite{Lorentz:1895}
erscheint. Enth"alt Beweis der Lorentzinvarianz der Gleichungen 
der Lorentzschen Elektrodynamik ("`Elektronentheorie"') in 
linearer Ordnung in $v/c$ und Besprechung des 
Michelson-Morley-Experiments samt der Erkl"arung des negativen 
Ausgangs durch Kontraktionshypothese. Dieses Werk 
war Einstein vor 1905 bekannt. 
\item
\emph{Einstein 1896:} Stellt sich die "`Aarauer Frage"': Was sehe ich, 
wenn ich einem Lichtstrahl mit Lichtgeschwindigkeit folge? Einstein 
bezeichnet dies als sein erstes Gedankenexperiment zur SRT.   
\item
\emph{Poincar\'e 1898:} Erkl"art, da"s wir keinen nat"urlichen 
gegebenen Begriff von Gleichzeitigkeit haben, sondern diesen 
durch Konventionen (etwa Signalaustausch) erst festlegen 
m"ussen~\cite{Poincare:1898}.    
\item
\emph{Lorentz 1899:}
Erstes Auftauchen des vollen analytischen Ausdrucks f"ur die 
Lorentz-Transformationen bei Lorentz~\cite{Lorentz:1899}. 
\item
\emph{Einstein:1899} 
Macht in Briefen an Mileva Mari\'c kritische Bemerkungen "uber 
die M"oglichkeit, dem "Ather einen Bewegungszustand zuzuschreiben.
Erw"ahnt ein selbst ausgedachtes Experiment, das den Einflu"s des 
"Athers auf die Lichtgeschwindigkeit in durchsichtigen K"orpern 
testen soll (wahrscheinlich "ahnlich dem von Fizeau), sowie eine 
"`Theorie "uber diese Sache"' (Details unbekannt).  
\item
\emph{Lorentz 1904:} Fast-Beweis der strikten Lorentzinvarianz 
seiner Elektronentheorie. Neue Zeitvariable $t'$, genannt 
"`Ortszeit"', wird nicht konsequent als physikalische Zeit 
behandelt. Gibt Ableitung der Geschwindigkeitsabh"angigkeit 
der tr"agen elektromagnetischen Masse und argumentiert, da"s 
diese generell f"ur alle tr"agen Massen g"ultig sein m"usse. 
Diese Arbeit \cite{Lorentz:1904} war Einstein 1905 nach eigener 
Aussage nicht bekannt.  
\item
\emph{Poincar\'e 1904:} "`Die gegenw"artige Krisis der mathematischen 
Physik."' Vortrag auf dem \emph{Congress of Arts and Science} in St.\,Louis
(USA), in dem er das Relativit"atsprinzip zu einem allgemeing"ultigen
Prinzip der Physik erhebt und eine Synchronisationsvorschrift durch 
Signalaustausch f"ur r"aumlich distante Uhren explizit schildert, 
die der Einsteinschen genau entspricht~\cite{Poincare:WdW}.     
\item
\emph{Poincar\'e 1905 
(Kurzversion~\cite{Poincare:1905}, Langversion~\cite{Poincare:1906}):}
Geht aus von der Unbeobachtbarkeit der Erdbewegung relativ zum 
"Ather (zitiert Experimente von Michelson und Morley).  
Erstes allgemeines Aussprechen eines mathematischen Invarianzprinzips 
f"ur die Elektrodynamik. Pr"agt die Ausdr"ucke  "`Lorentz-Transformation"' 
und "`Lorentzgruppe"'. Zeigt auch als erster, da"s Lorentz-Transformationen 
zusammen mit Drehungen eine Gruppe bilden. Bleibt aber im Formalen und 
bringt $t'$ nicht in Verbindung mit tats"achlicher 
Zeitmessung. 
\item
\emph{Einstein 1905 (eingereicht am 30. Juni)~\cite{Einstein:SRT}:}
Originalarbeit zur SRT. Geht aus von 1)~allgemeiner G"ultigkeit 
des Relativit"atsprinzips, 2)~Konstanz der Lichtgeschwindigkeit 
und 3) einer in allen Inertialsystemen gleichlautenden Definition 
der Gleichzeitigkeit r"aumlich distanter Ereignisse. Leitet damit 
Lorentz-Transformationen ab und gibt somit als erster eine 
physikalisch-operationale Deutung der Lorentz-Kontraktion und 
Zeitdilatation auf kinematischer Ebene (d.h. als Resultat einer 
allgemeinen Invarianzforderung dynamischer Gesetze).  
\end{itemize}

\end{document}